\documentclass[3p]{elsarticle}

\usepackage{lineno}
\usepackage{subfigure}
\usepackage{verbatim}
\usepackage{amsmath}
\usepackage{amsfonts}
\usepackage{amssymb}
\usepackage{graphicx}
\usepackage{lscape}
\usepackage{multicol}
\usepackage{centernot}
\usepackage{floatpag}

\usepackage{adjustbox}
\usepackage{array}
\usepackage{booktabs}
\usepackage{multirow}

\usepackage{colortbl}
\definecolor{Grey9}{rgb}{0.9,0.9,0.9}
\definecolor{Grey7}{rgb}{0.7,0.7,0.7}
\newcolumntype{R}[2]{%
    >{\adjustbox{angle=#1,lap=\width-(#2)}\bgroup}%
    l%
    <{\egroup}%
}
\newcommand*\rot{\multicolumn{1}{R{90}{0.1pt}}}

\journal{Computer Science Review}

\begin{document}

\begin{frontmatter}

\title{Scheduling in distributed systems: a cloud computing perspective}


\author[unicamp]{Luiz F. Bittencourt}
\ead{bit@ic.unicamp.br}

\author[usp]{Alfredo Goldman}
\ead{gold@ime.usp.br}

\author[unicamp]{Edmundo R. M. Madeira}
\ead{edmundo@ic.unicamp.br}

\author[unicamp]{Nelson L. S. da Fonseca}
\ead{nfonseca@ic.unicamp.br}

\author[manchester]{Rizos Sakellariou}
\ead{rizos@cs.man.ac.uk}

\address[unicamp]{Institute of Computing, University of Campinas, Brazil}
\address[usp]{Instituto de Matem\'{a}tica e Estat\'{i}stica, Universidade de S\~{a}o Paulo, Brazil}
\address[manchester]{School of Computer Science, University of Manchester, UK}

\begin{abstract}
Scheduling is essentially a decision-making process that enables resource sharing among a number of activities by determining their execution order on the set of available resources. The emergence of distributed systems brought new challenges on scheduling in computer systems, including clusters, grids, and more recently clouds. On the other hand, the plethora of research makes it hard for both newcomers researchers to understand the relationship among different scheduling problems and strategies proposed in the literature, which hampers the identification of new and relevant research avenues. In this paper we introduce a classification of the scheduling problem in distributed systems by presenting a taxonomy that incorporates recent developments, especially those in cloud computing. We review the scheduling literature to corroborate the taxonomy and analyze the interest in different branches of the proposed taxonomy. Finally, we identify relevant future directions in scheduling for distributed systems.
\end{abstract}


\end{frontmatter}



\section{Introduction} The scheduling problem arises in countless areas, and it evolves over time along with industry and technology~\cite{PinedoScheduling}. 	With the development of computers, scheduling in computer processors received great attention~\cite{DCP,KwokStatic1999}, being the most common objective the minimization of  task completion times, also known as makespan. Besides some peculiarities, the basic principles remained the same as in scheduling activities among machines in production. In supercomputers, multiprocessor scheduling considers several parallel processors with the same capacity. In addition, the data source is considered to be centralized and connected by a high speed channel between processors, in a way that activities (or jobs) can exchange messages quickly.
 	
	More recently, computer networks allowed clusters of homogeneous computers to act as a multiprocessor computer with distributed data sources. However, when compared to supercomputers, clusters initially had a slow communication channel between processors, which made data exchange among processors more expensive. The scheduling of jobs in computing clusters led to another branch of research: the scheduling in distributed computer systems. With improvements in computer networks, the connection among computing nodes in clusters became faster. On the other hand, new applications demanded more and more bandwidth, storing and exchanging massive volumes of data. Multimedia and e-Science are examples of applications that handle large data sets nowadays, putting in evidence the importance of communications  to improve performance and support quality of service offering in distributed systems.

	Grid computing emerged in the late 90's as a heterogeneous collaborative distributed system~\cite{FosterAnatomy} evolved from homogeneous distributed computing platforms. Grids are shared systems that enclose potentially any computing device connected to a network, from workstations to clusters. Computing grids are infrastructures that enable resource sharing by establishing use policies as well as security rules, which compose the so called Virtual Organizations (VOs)~\cite{FosterAnatomy}.

	Cloud computing offers computing resources, often virtualized, as services to the users, hiding technical aspects regarding resource management~\cite{Bittencourt-WileyChapter2015}. Therefore, clusters and grids can be part of datacenters in the cloud computing infrastructure, demanding new optimization objectives and variables common in green computing~\cite{GreenComputing} and utility computing~\cite{UtilityComputing}.

	Kwok and Ahmad stated in 1999~\cite{KwokStatic1999} that considering heterogeneous platforms was a challenging direction to extend scheduling algorithms. As a consequence of the popularization of these platforms in grid and cloud computing, novel scheduling concepts appeared in the iterature~\cite{HEFT,Lucchese2006,JCCPE,HamscherEvaluation2000,YuBudget}. While on the one hand fundamental scheduling aspects remain unchanged, on the other hand different optimization objectives ballooned the scheduling literature in the past decade. Such swell in the field brought so much information that it became challenging the recognition of the exact contribution of new results results. Since challenges in scheduling still exist, Smith argues in~\cite{StephenSolved} that scheduling is not a fully solved problem, he stated that \textit{``Scheduling techniques that properly account for uncertainty, enable controlled solution change, and support efficient negotiation and refinement of constraints are crucial prerequisites, and the need to operate in the context of multiple self-interested agents is a given.''}. This statement matches certain characteristics in virtualization and service  in  cloud computing, as we shall describe in the upcoming sections.
	
	In a nutshell, this paper has three main contributions:
	\begin{enumerate}
		\item Propose a taxonomy for scheduling in distributed systems and introduce a taxonomy extension to cover cloud computing schedulers.
		\item Classify the literature in the proposed taxonomy;
		\item Identify relevant future directions for scheduling in distributed systems.
	\end{enumerate}
	
	Due to its wide application, there exist a variety of approaches to the scheduling problem. This paper presents directives for scheduling researchers to identify and classify their work, as well as to provide them with an overview of existing approaches associated to their research by presenting a broad view of the scheduling problem in distributed systems as well as introducing existing works. We first introduce the problem of scheduling in distributed systems, covering  advances in scheduling in cluster and grid computing. Then, an overview of the proposed taxonomy is presented, followed by the state-of-the-art in each branch of the taxonomy tree. After that, we focus on cloud computing and detail similarities and differences of scheduling in clouds with scheduling in previously existing distributed systems. We introduce a taxonomy of scheduling in cloud computing, extending the pre-cloud taxonomy. Finally, we discuss research challenges that were inherited from grid and cluster computing by the cloud computing paradigm, as well as new problems to be tackled.

 This paper is organized as follows. Section~\ref{sec:rel} discusses previous work that have addressed reviews and surveys of scheduling in distributed computing. Section~\ref{sec:concepts} introduces basic concepts to define scheduling problem. Section \ref{sec:overview} briefly introduces the whole taxonomy discussed in this paper, highlighting branches that were introduced to cover scheduling in cloud computing. Section~\ref{sec:pre-taxonomy} presents a taxonomy of schedulers previously to the advent of cloud computing (\textit{pre-cloud} taxonomy). Section~\ref{sec:clouds} discusses and classifies the scheduling taxonomy in cloud computing (\textit{cloud taxonomy}). Section~\ref{sec:survey} reviews  the cloud computing scheduling literature according to the proposed taxonomy. Future directions in scheduling for distributed systems are discussed in Section~\ref{sec:chal}, and Section~\ref{sec:con} presents the concluding remarks.

\section{Related Work}
\label{sec:rel}	
	In computer science, with the constant networking and middleware development, scheduling in distributed processing systems is one of the topics which has gained attention in the last two decades. Casavant and Kuhl~\cite{CasavantTaxonomy1988} present a taxonomy of scheduling in general purpose distributed systems. The classification presented by the authors include local and global, static and dynamic, distributed and non-distributed, cooperative and non-cooperative scheduling, as well as some approaches to solve the problem, such as optimal and sub-optimal, heuristic, and approximate. This presented classification is complete in some sense, and it is still valid nowadays. However the current state of distributed systems indeed  demands the addition of new branches in this taxonomy.

	Kwok and Ahmad~\cite{KwokStatic1999} survey static scheduling algorithms for allocating tasks connected as directed task graphs (DAGs) into multiprocessors. The authors presented a simplified taxonomy for approaches to the problem, as well as the description and classification of $27$ scheduling algorithms. The DAG scheduling algorithms for multiprocessors have been adapted for scheduling in distributed systems, incorporating intrinsic characteristics of such systems for an enhanced performance. Therefore, Kwok and Ahmad presented static scheduling algorithms for multiprocessors, which are also applicable to distributed systems, and their classification. In this paper we review extensions of those algorithms as well as the their classification by including heterogeneous systems, dynamic scheduling algorithms, scheduling algorithms in modern distributed environments, and new scheduling techniques.

	In the last decades, after Kwok and Ahmad's work, other surveys and taxonomies for solutions to the scheduling problem for parallel systems have been developed. Most of these works focus on heterogeneous distributed systems~\cite{JiangSurveyGrids2007}, which Ahmad and Kwok considered as one of the most challenging directions to follow~\cite{KwokStatic1999}.
	
	Job scheduling strategies for grid computing are evaluated by Hamscher et al. in~\cite{HamscherEvaluation2000}. The authors present common scheduling structures, such as centralized, decentralized, and hierarchical. Within each scheduling structure, they present and evaluate $4$ processor selection strategies and three scheduling algorithms, namely \emph{First-Come-First-Serve} (FCFS), \emph{Random}, and \emph{Backfill}. After this work, many dynamic scheduling strategies were developed to tackle with the grid dynamicity.

	As Feitelson et al. claim in their scheduling review \cite{FeitelsonStatusReport}, parallel job scheduling reviews are needed in a regular basis. The purpose of their short review was to introduce clusters and grids into the parallel job scheduling literature. Indeed, the authors present an introduction to job scheduling in grids, highlighting differences between a parallel computer and the grid. They point out cross-domain load balancing and co-allocations as two main concerns when scheduling in grids. In our work, we introduce a classification of schedulers in distributed systems that comprises a more extensive view of grid computing algorithms. Moreover, we highlight new requirements for the cloud computing emergent paradigm as well as its differences to grid computing.

	Wieczorek et al.~\cite{WieczorekMulticriteriaTaxonomy2008} present a taxonomy in the scheduling problem for workflows considering multiple criteria optimization in grid computing environments. The authors separate the multi-criteria scheduling taxonomy in $5$ \emph{facets}, namely \emph{scheduling process, scheduling criteria, resource model, task model}, and \emph{workflow model}, each facet describing the problem from a different point of view. These facets are expanded to classify existing works in a smaller granularity, pointing out where current research can be expanded and the work in each facet. 

	We highlight two conclusions achieved by the authors in ~\cite{WieczorekMulticriteriaTaxonomy2008} which are touched by contributions given by our survey: (i) ``grid workflow scheduling problem is still not fully addressed by existing work''; and (ii) ``there are almost no workflow scheduling approaches which are based on an adaptive cost model for criteria''. As a contribution to (i), in this survey we expand the general distributed system scheduling to comprise it. As a contribution to (ii), we include scheduling taxonomies for utility grids and cloud computing environments.

	Besides specific reviews of the scheduling literature, some other general surveys include scheduling taxonomies and classification. In~\cite{YuTaxonomyJGC2005}, Yu and Buyya classify workflow management systems for grid computing. The presented classification includes some taxonomy branches that have influence in schedulers, such as the workflow structure (DAG or non-DAG), workflow QoS constraints, and information retrieval coordination. Regarding the scheduling itself, the classification includes system \textit{architecture, decision making, planning scheme, strategies, and performance estimation}~\cite{YuTaxonomyJGC2005}. In this paper we cover this classification and further expand it to a more general view of the scheduling problem, comprising independent tasks and recent advances in distributed systems, such as cloud computing. Venugopal et al. review the literature on data grids in~\cite{VenugopalTaxonomy2006}. Concerning scheduling in such systems, the authors present a brief taxonomy that includes classifications regarding \textit{application model, scope, data replication, utility function, and locality}~\cite{VenugopalTaxonomy2006}.

	In the last decade, cloud computing taxonomies started to appear.  In~\cite{hilley2009cloud}, Hilley describes a bunch of existing services that provide cloud computing infrastructures, classifying them inside a proposed taxonomy.  Oliveira et al.~\cite{OliveiraTaxonomy2010} classify the cloud computing paradigm regarding many aspects, such as architecture, pricing, privacy, and technology used.  Other recent review papers in the scheduling literature have focused on specificities of scheduling, as for example meta-heuristics~\cite{Wei2014}, scheduling techniques~\cite{Shaw2014}, workflow costs~\cite{Alkhanak2015}, virtual machines~\cite{Pietri:2016:MVM:2988524.2983575}, workflows in clouds~\cite{Fuhui2015}, evolutionary approaches, \cite{Zhan2015}. In this paper we introduce more general cloud computing classification aspects that affect the development of scheduling algorithms, which are not addressed by those papers.
	
	Scheduling has been a focus of research with the high attention given by the community to grid computing in the last decade and to cloud computing in the last few years. With this, algorithms and techniques were freely developed, aggregating new concepts from the service oriented computing and aspects from the constantly changing distributed systems environments. As shown in this section, many advances were achieved in the last decades. Also, excellent reviews which unify topics surveyed in this paper exist, however, updates are necessary in subtopics such as grid and cloud computing. In this work, we classify new developments in the scheduling field in what concerns the distributed computing systems evolution, aiming at fitting topics developed in the last decade and future directions. This paper brings an organized overview of the scheduling problem advancements in distributed computer systems, serving as a reference for the development of algorithms for cloud computing and the upcoming distributed computing paradigms.

\section{Basic Concepts}
\label{sec:concepts}
	The definition of the scheduling problem given by Pinedo in~\cite{PinedoScheduling} is as follows:\\
	
	``\emph{Scheduling is a decision-making process that is used on a regular basis in many manufacturing and services industries. It deals with the allocation of resources to tasks over given time periods and its goal is to optimize one or more objectives.}''\\
	
	There exist many other definitions to the general scheduling problem in the literature~\cite{Ha89}. The scheduling problem is NP-Complete, although there exist polynomial time solutions for few scenarios~\cite{PinedoScheduling}. 
	
	In a distributed system, the scheduler has the role of choosing in which computational resource each task (or job) will execute, therefore distributing jobs to be executed concurrently. The execution time of each job, and consequently  the completion time of all jobs, is intrinsic to the generated schedule. The scheduler organizes jobs in resources queues usually following an objective function. The most common objective found in the distributed systems literature is minimize the completion time, or makespan, achieved by proper allocation of tasks in the available resources. Common objective functions are examined in more details in Section~\ref{sec:objectives}.

\subsection{Scheduling Model}
	Here we describe a scheduling model that is the basis to the scheduling problems in distributed systems, as previously modeled in~\cite{BittencourtJISA}. Consider a distributed system composed of a set of computational resources $\mathcal{R}=\{r_1, r_2, ..., r_k\}$, with associated processing capacities $p_{r_i} \in \mathbb{R}^+$, connected by network links. Each resource $r_i$ has a set of links $\mathcal{L}_i=\{l_{i,1}, l_{i,2}, ... , l_{i,m}\}$, $1 \leq m \leq k$, where $l_{i,j} \in \mathbb{R}^+$ is the available bandwidth in the link between resources $r_i$ and $r_j$, with $l_{i,i}=\infty$.  Let $\mathcal{A}=\{a_1,a_2,...,a_n\}$ be the set of application jobs submitted by users and to be executed in $\mathcal{R}$. The scheduler is responsible for assigning a sequencing of jobs in $\mathcal{A}$ onto the resources in $\mathcal{R}$ according to one or more objectives.
	
	According to the well-known, general scheduling problem description by Pinedo~\cite{PinedoScheduling}, the scheduling problem is commonly described as a triplet $\alpha~|~\beta~|~\gamma$. The $\alpha$ field, with a single entry, is a description of the execution environment. The $\beta$ field can have from zero to multiple entries, and it describes what is the processing to be done. Lastly, the $\gamma$ field entry is the description of the objective to be optimized. In this paper we are mainly interested in discussing the scheduling problem for two different $\alpha$'s~\cite{PinedoScheduling}: 
	
	\begin{enumerate}
		\item $~\alpha=P_m$: Identical machines in parallel. Execution environment is composed of $m$ identical machines in parallel, as usually found in a cluster computing infrastructure.
		\item $~\alpha=R_m$: Unrelated machines in parallel. There are $m$ different machines in parallel, and machine $i$ executes job $j$ at speed $v_{i,j}$, as in a grid or cloud computing infrastructure. Note that, given two application jobs $a_1$ and $a_2$ and two computational resources $r_1$ and $r_2$, $v_{a_1,r_1} > v_{a_2, r_1} \centernot\implies v_{a_1,r_2} > v_{a_2, r_2}$.
	\end{enumerate}
	
	As utility computing becomes more popular, we propose the introduction of a new field to the representation of the scheduling problem: $\alpha~|~\beta~|~\gamma~|~\mu$, where the $\mu$ field describes the charging models adopted in the system. For example, $\mu$ can be an hourly-based charge, an auction system, per-transaction charge, data-access charge, and so on. This is well-suited to the cloud computing paradigm, in which many different computing services can be offered with a variety of charging models, as well as for the upcoming utility-based computing and networking infrastructures.

\subsection{System Organization}	
\begin{figure*}[!htbp]
\centering
\subfigure[Centralized infrastructure.]
{
        \includegraphics[width=0.42\textwidth]{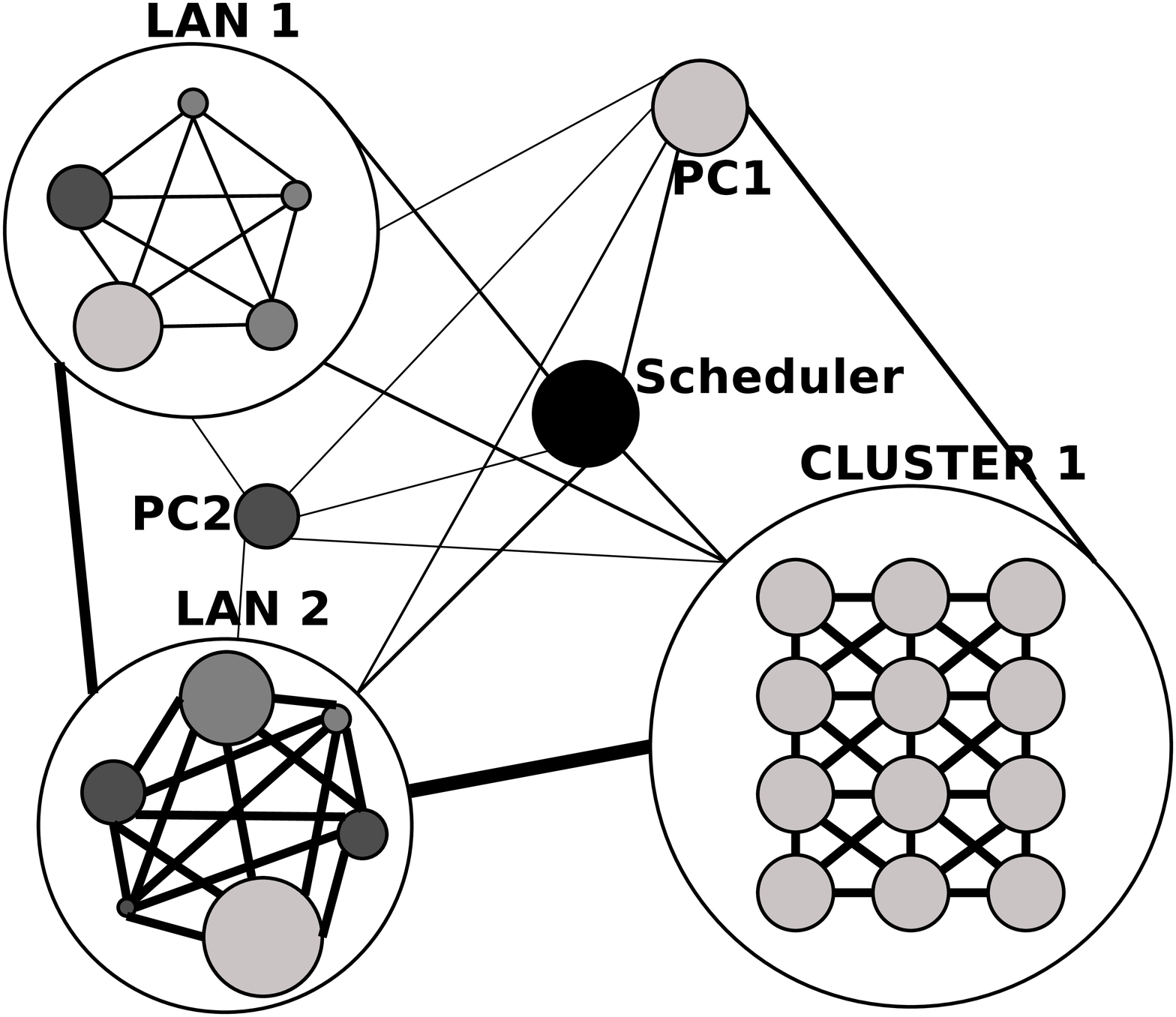}
	\label{fig:infraCentral}
}
\hspace{1.0cm}
\subfigure[Decentralized infrastructure.]
{
        \includegraphics[width=0.35\textwidth]{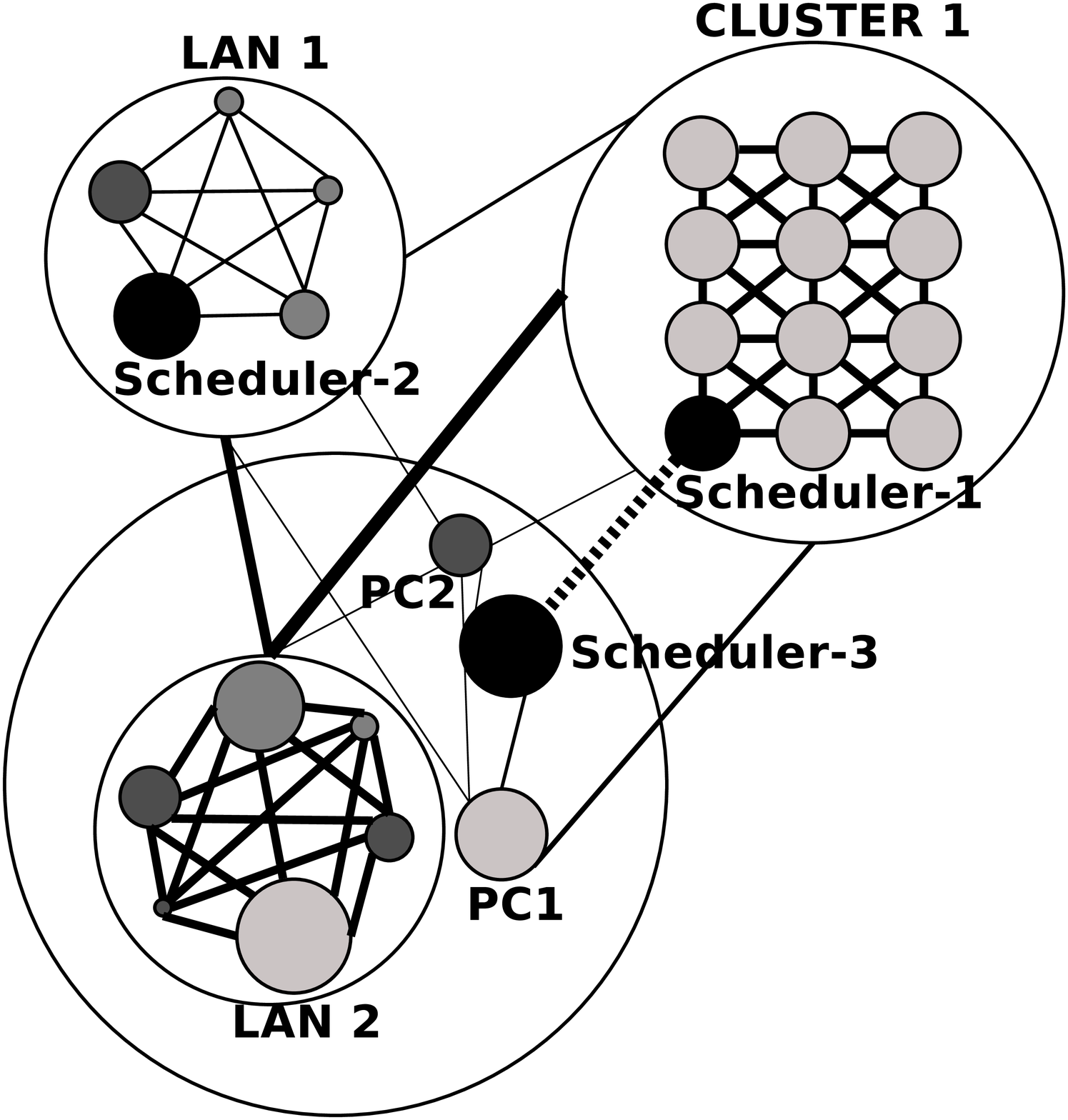}
	\label{fig:infraDecentral}
}

\subfigure[Hierarchical infrastructure.]
{
	\includegraphics[width=0.9\textwidth]{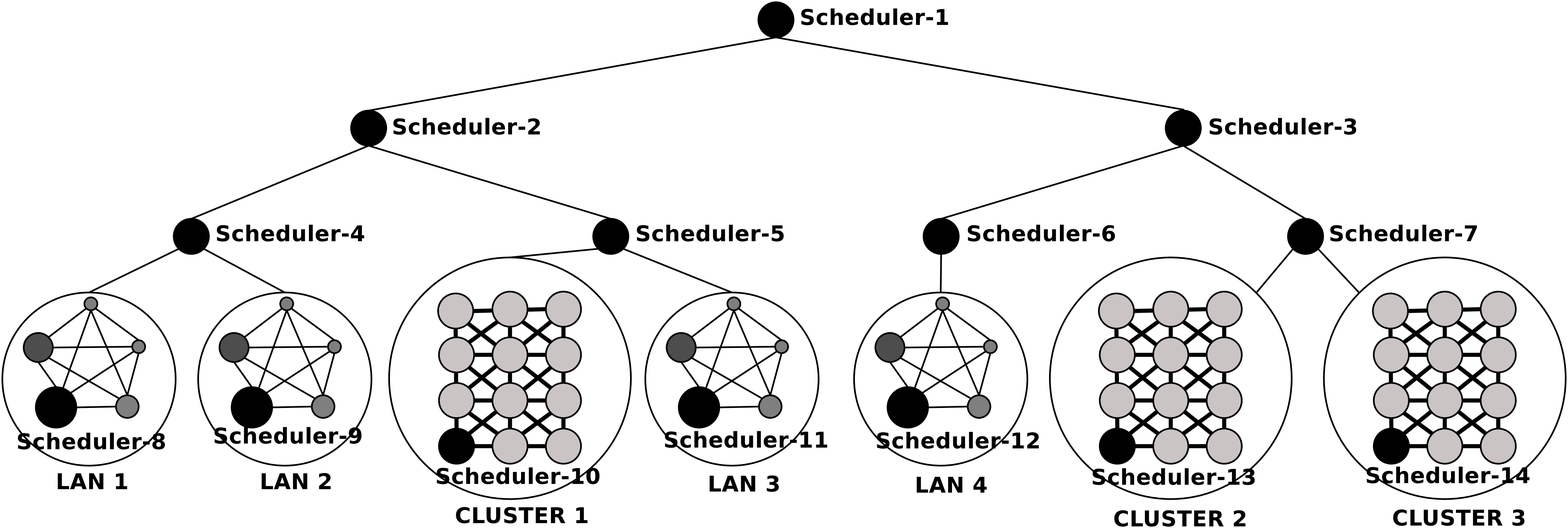}
	\label{fig:infraHierarchical}
}
\caption{Architectural organization of distributed systems and schedulers.}
\label{fig:organization}
\end{figure*}

	In distributed systems, the overall system organization may have influence in the scheduler scope. This organization can be a result of the infra-structure architecture or existing policies and access controls. Regarding the system management organization, it can be classified in three categories: \emph{centralized}, \emph{decentralized}, and \emph{hierarchical} (Figure~\ref{fig:organization}). 
	
	In a \emph{centralized} management, the distributed system has a central entity that has information about the whole system and is responsible for managing its resources and jobs execution. On the other hand, a \emph{decentralized} management is accomplished through distributed algorithms that manage the system without a central entity. In such systems, management roles may be distributed over resources in the system~\cite{KlausTaxonomy}. Scheduling in a decentralized system is made by more than one scheduler, where each scheduler has information about part of the system and may exchange information with other schedulers to make decisions in a cooperative way. A decentralized non-cooperative distributed system considers that each scheduler is responsible for achieving its own objectives, regardless the objectives of other schedulers.  A \emph{hierarchical} system structures the management in a tree-like structure. Therefore, machines at a given level can communicate with machines at a level immediately above or below it in the hierarchy~\cite{KlausTaxonomy}. With this, the distributed system management is shared, with entities that are higher in the hierarchy taking high-level decisions (e.g., in which WAN a job will execute), while entities at the bottom make the low-level decisions (e.g., in which machine a job will be executed).

	Note that the centralized architecture can be seen as a specific case of the decentralized model where only one scheduler exists. In addition, algorithms for the centralized architecture can independently execute in each scheduler of a decentralized (or hierarchical) system. In this case, a scheduler component could be responsible for the communication among different schedulers to arrange jobs execution within different boundaries. A more detailed view of distributed systems organization can be found in~\cite{VenugopalTaxonomy2006}. 

\begin{figure}[!htbp]
\centering
	 \includegraphics[width=0.55\textwidth]{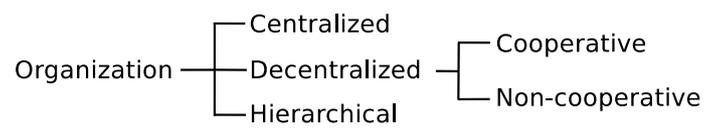}
\caption{Taxonomy for schedulers organization in distributed systems.}
\label{fig:tax_Organization}
\end{figure}

\section{Taxonomy overview}
\label{sec:overview}
\begin{figure}[!htbp]
	\centering
	 \includegraphics[height=0.97\textheight]{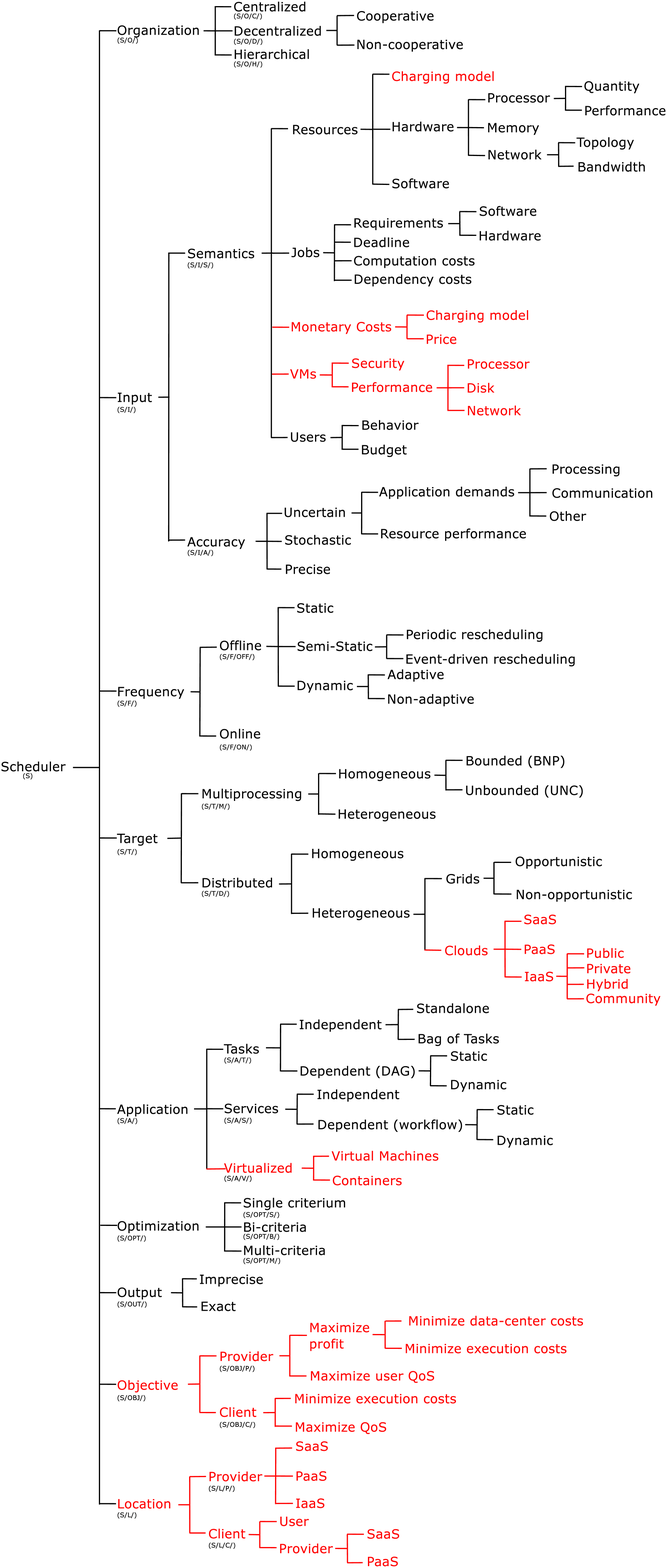}
	\caption{Overview of the proposed taxonomy. A short text notation is shown below branches up to the third level, which can be extended to deeper branches in the taxonomy.}
	\label{fig:tax_full}
\end{figure}

	This section introduces the taxonomy presented in this paper without discussing details. Figure \ref{fig:tax_full} presents the full taxonomy, where each branch up to the third level presents a short text notation, which can be extended to deeper branches in the taxonomy, and that is used later in this paper to classify the current literature. 
	
	The taxonomy brings a classification according to different factors derived from schedulers developed for distributed systems that existed before the advent of cloud computing, which we call \textit{pre-cloud schedulers}, and those developed for cloud computing infrastructures, the \textit{cloud schedulers}. For pre-cloud schedulers, taxonomy branches are the scheduler \textbf{\emph{organization}}, as discussed in the previous section, \textbf{\emph{input}, \emph{frequency}, \emph{target}, \emph{application}, \emph{optimization}},  and \textbf{\emph{output}}. For cloud schedulers, we update the taxonomy (branches in red in Figure \ref{fig:tax_full}) to reflect characteristics that were introduced to address cloud computing scheduling, namely branches about \textbf{\emph{monetary costs}} and \textbf{\emph{virtual machines}} under the \emph{input} branch, \textbf{\emph{clouds}} under the \emph{target} branch, \textbf{\emph{virtualized}} application models, scheduler \textbf{\emph{objective}}, and \textbf{\emph{location}}. Detailed discussion on each branch of the taxonomy is presented in the upcoming sections.

\section{Pre-cloud scheduling taxonomy}
\label{sec:pre-taxonomy}

\subsection{Input Classification}
\begin{figure}[!htbp]
	\centering
	 \includegraphics[width=0.2\textwidth]{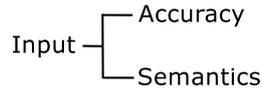}
	\caption{Input taxonomy.}
	\label{fig:tax_InputData}
\end{figure}

	The information available to the scheduler drives the decision making process. This input information is subject to variations in \emph{accuracy} and \emph{semantics} (different characteristics influencing the scheduling process), as shown in Figure~\ref{fig:tax_InputData}.

\subsubsection{Input Accuracy}
\begin{figure}[!htbp]
	\centering
	 \includegraphics[width=0.65\textwidth]{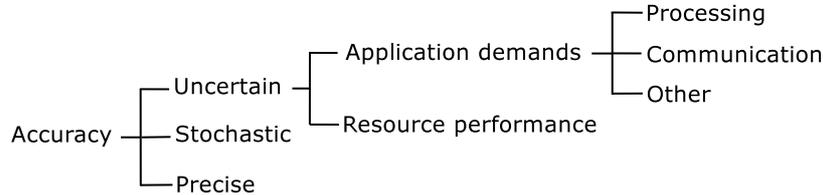}
	\caption{Input accuracy taxonomy.}
	\label{fig:tax_InputAccuracy}
\end{figure}

	The scheduling algorithm may expect the input data to be \emph{precise}, \emph{uncertain}, or \emph{stochastic}. When the input data is considered to be \emph{precise}, the algorithm is developed to work better within conditions in which both the input data about jobs and about resources reflect the real execution behavior. 	Input data is said to be stochastic when it is expected to follow some generalized probability distribution. If the input data is assumed to be \emph{uncertain}, the algorithm can make use of mechanisms to deal with different kinds of uncertainty, including \textit{application demands} and \textit{resources availability} uncertainty. Schedulers commonly assume that application communication and computation costs are known. Therefore, we consider that application specification can present uncertainty regarding its \textit{communication} and \textit{computation (processing)} demands, and also \emph{other} demands as further possible application requirements (memory, storage, etc.). The input accuracy taxonomy is shown in Figure~\ref{fig:tax_InputAccuracy}.
	
	Derbal presents in~\cite{Entropic} an entropy-based quantification of input data uncertainty regarding resources performance. Batista and Fonseca deal with both application and resource availability uncertainties in~\cite{Batista2011-1}. Examples of studies in stochastic scheduling are Schopf~\cite{Schopf1999} and earlier by Pinedo~\cite{Pinedo1983}. Precise input data is commonly found in the literature, such as in works by Topcuoglu et al.~\cite{HEFT} and Kwok et al.~\cite{KwokStatic1999}.

\subsubsection{Input Semantics}
\begin{figure}[!htbp]
	\centering
	 \includegraphics[width=0.65\textwidth]{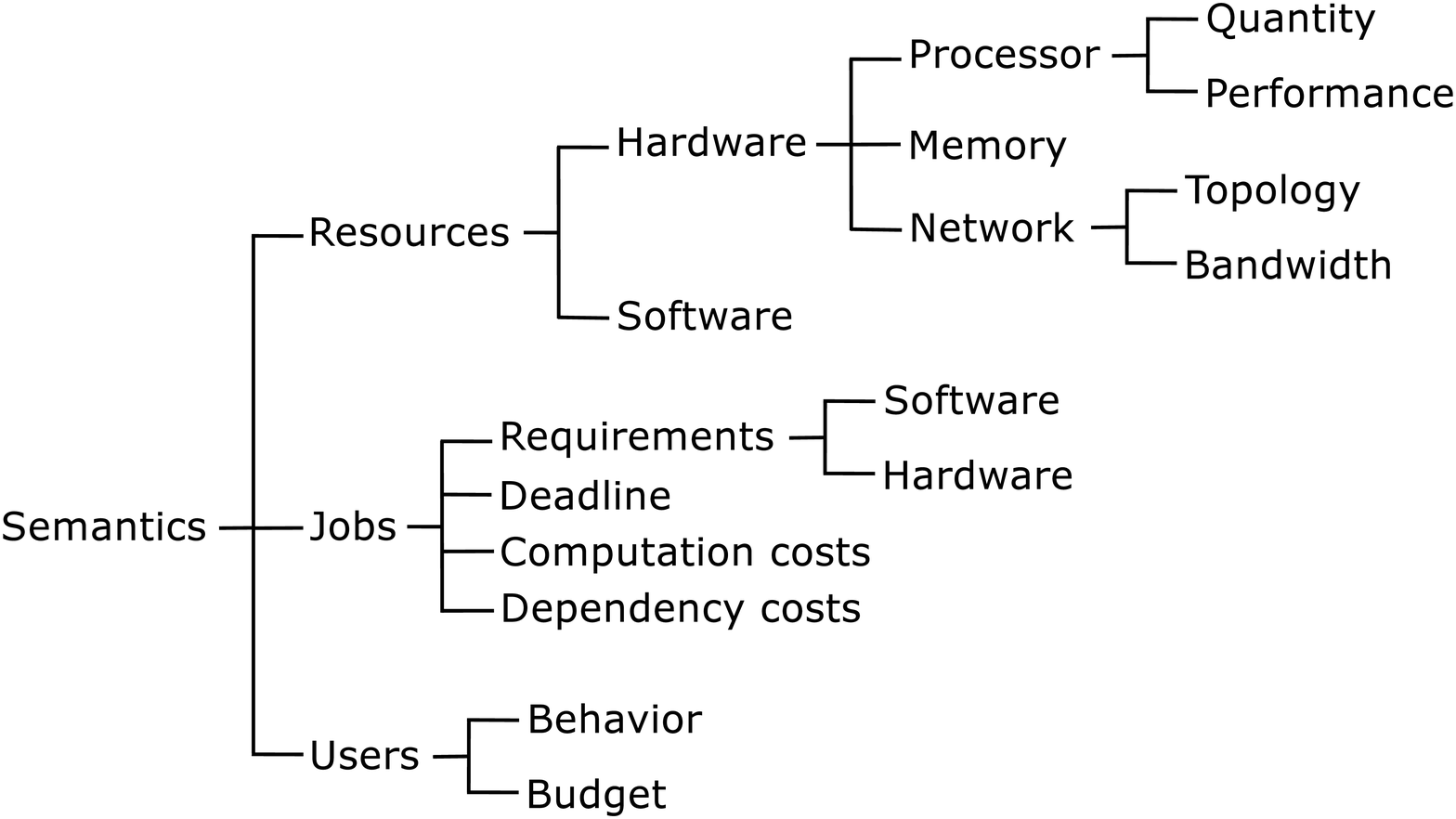}
	\caption{Input semantics taxonomy.}
	\label{fig:tax_InputSemantics}
\end{figure}

	The input data semantics taxonomy (Figure~\ref{fig:tax_InputSemantics}) classifies the \textit{meaning} of the information provided to the scheduling algorithm. This information is used by the scheduler to perform the decision making following pre-defined objectives. The semantics of the input data can be related to \emph{resources}, \emph{jobs}, or \emph{users}. Information about \emph{resources} are divided into \emph{software} and \emph{hardware}. Hardware information commonly comprises \emph{processor}, \emph{memory}, and \emph{bandwidth}. The \emph{quantity} and \emph{performance} of processors are usually considered as input data. Within the \emph{software} branch below \emph{resources}, applications, services, and libraries  available on each resource can be provided to the scheduler. 
	
	A job may have \emph{hardware} and \emph{software} requirements. Therefore, the scheduler can match software and hardware requirements from jobs with the available hardware and software on each resource. A job may also have a \emph{deadline}, that is, a maximum completion time. An important information used by the scheduler is the \emph{computation cost} of a job, which influences the decision on where it will be executed. Another important information for coupled jobs are their \emph{dependencies costs}, which influences where dependent jobs should be placed in order to minimize their data transmission times or any other objective.
	
	The \emph{user} who submits a job may have a \emph{budget} to follow. This information is usual when the target system is a utility grid or a cloud and the scheduler must use it to check if a given schedule obeys this user's constraint. Another information regarding \emph{users} is their \textit{behavior}. This can be useful in a shared system (such as grids), in which the scheduler may use information about which applications are most often executed by a user $A$, therefore estimating how long they take to run and how much $CPU$ they consume. For example, a scheduler may expect that a job from user $A$ running for $30$ minutes is about to finish, since user $A$ jobs usually do not take longer than half an hour.
		
	The execution time of a job is, in general, highly dependent on the processor on which it runs. Because of that, scheduling algorithms usually consider the processing capacities as an important, and often determinant, characteristic 
when deciding where each job will execute. Research assuming that the scheduler has information about resources bandwidth, memory, and processors quantity and performance, as well as jobs computation and communication costs, is very common~\cite{Pandey2010, Hagras2003, Karatza03, KwokSemi-Static, denBossche2010}. Information about software installed in the resources is also considered in works where jobs have software restrictions~\cite{schedulingGrads}. Ramamrithan et al. present a set of heuristics to schedule jobs with deadlines and resources requirements in~\cite{Ramamrithan1989}. Chaves et al. present an integer programming formulation for scheduling grid applications with software requirements in~\cite{Chaves2011}. In~\cite{denBossche2010}, den Bossche et al. deal with the problem of scheduling deadline-constrained workloads, thus considering jobs deadlines as input data. Yu and Buyya in~\cite{YuBudget} consider the user's budget when scheduling workflows in grids. In \cite{FoxGGF}, Fox and Gannon overviews contributions on workflow in grid systems. In~\cite{Gomes2011}, Gomes and Costa analyze desktop users' application behavior in opportunistic grids, extracting the local application CPU usage to predict bursts durations, creating profiles to be used in scheduling decisions.

\subsection{Frequency}
	The scheduling frequency is associated to how many times the scheduler is executed within a period of time. In a general scope, the scheduling algorithm can be classified as \emph{offline} or \emph{online}. An online algorithm assumes that its input is a data sequence of unknown size, handling parts of the data without having the whole input at the start of its execution. Therefore, an online scheduling algorithm does not know the entire set of jobs being scheduled at a given time (or even the whole set of resources where the jobs will execute). Rather, it schedules a job without information about the next incoming jobs. Thus, the decision maker becomes aware of the existence of a job only when the job is released and presented to it~\cite{PinedoScheduling}. An online scheduling algorithm can be aided by some information about the input stream, such as which probability distribution the input should follow. On the other hand, in the more common paradigm of \emph{offline scheduling}, the information regarding all jobs is known \emph{a priori}. Note that the online paradigm can have an intersection with the offline paradigm if arriving jobs are grouped into sets before they are scheduled, in a \emph{batch-mode} scheduling~\cite{Maheswaran99}.
	
\begin{figure}[!htbp]
	\centering
	 \includegraphics[width=0.65\textwidth]{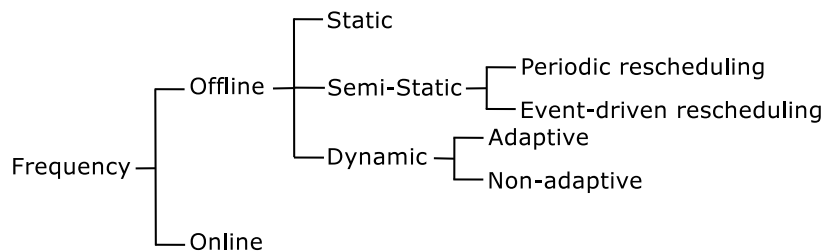}
	\caption{Taxonomy according to the scheduling frequency.}
	\label{fig:tax_schedFrequency}
\end{figure}

	In another branch, offline scheduling can be classified as \emph{static} or \emph{dynamic}. Static schedulers perform scheduling at \emph{compile time}. This means that the scheduling algorithm decision making is accomplished using information available before any job is sent to the resources execution queues~\cite{KwokStatic1999}, and no interference is made until the end of the application execution. Conversely, a dynamic scheduler uses information which is updated during jobs execution, thus having an up-to-date prospect of the system as a whole. To do this, the dynamic scheduler dispatches a set of jobs to execution and observes the behavior of the system, gathering information to schedule the next set of jobs. This information gathering is usually done when a resource becomes idle~\cite{Lucchese2006}. As extensions from these two general classifications, \emph{semi-static} (also called \emph{hybrid}~\cite{Maheswaran99}) and \emph{adaptive} approaches exist as well~\cite{KwokSemi-Static,JCCPE}. In a semi-static approach, an initial mapping is done and re-mappings are performed when necessary. This rescheduling can be done \textit{periodically} or after observing some \textit{event}, such as resource performance changes during execution or application input data changes~\cite{KwokSemi-Static}. Adaptive approaches alter the dynamicity according to the observed behavior of the system, therefore being able to act as a highly dynamic algorithm, which updates the system information frequently, or even as an almost static one. The taxonomy regarding the scheduling frequency is presented in Figure~\ref{fig:tax_schedFrequency}.
			
	Offline algorithms are most commonly found in the literature than online algorithms. Examples of well-known static offline scheduling algorithms are the \emph{Heterogeneous Earliest Finish Time} (HEFT)~\cite{HEFT} and the \emph{Dynamic Critical Path} (DCP)~\cite{DCP}. A semi-static algorithm with periodic rescheduling is presented by Kwok et al. in~\cite{KwokSemi-Static}. Lucchese et al. present a semi-static algorithm with event-driven rescheduling in~\cite{Lucchese2006}. A round-based adaptive dynamic algorithm is presented by Bittencourt and Madeira in~\cite{JCCPE}. Maheswaran et al. present non-adaptive dynamic approaches in~\cite{Maheswaran99}, as well as online scheduling algorithms. Other examples of online algorithms include~\cite{CastilloOnlineGrids,SchwiegelshohnOnlineGrids,DynamicOnline}.

\subsection{Target System}
\begin{figure}[!htbp]
	\centering
	 \includegraphics[width=0.65\textwidth]{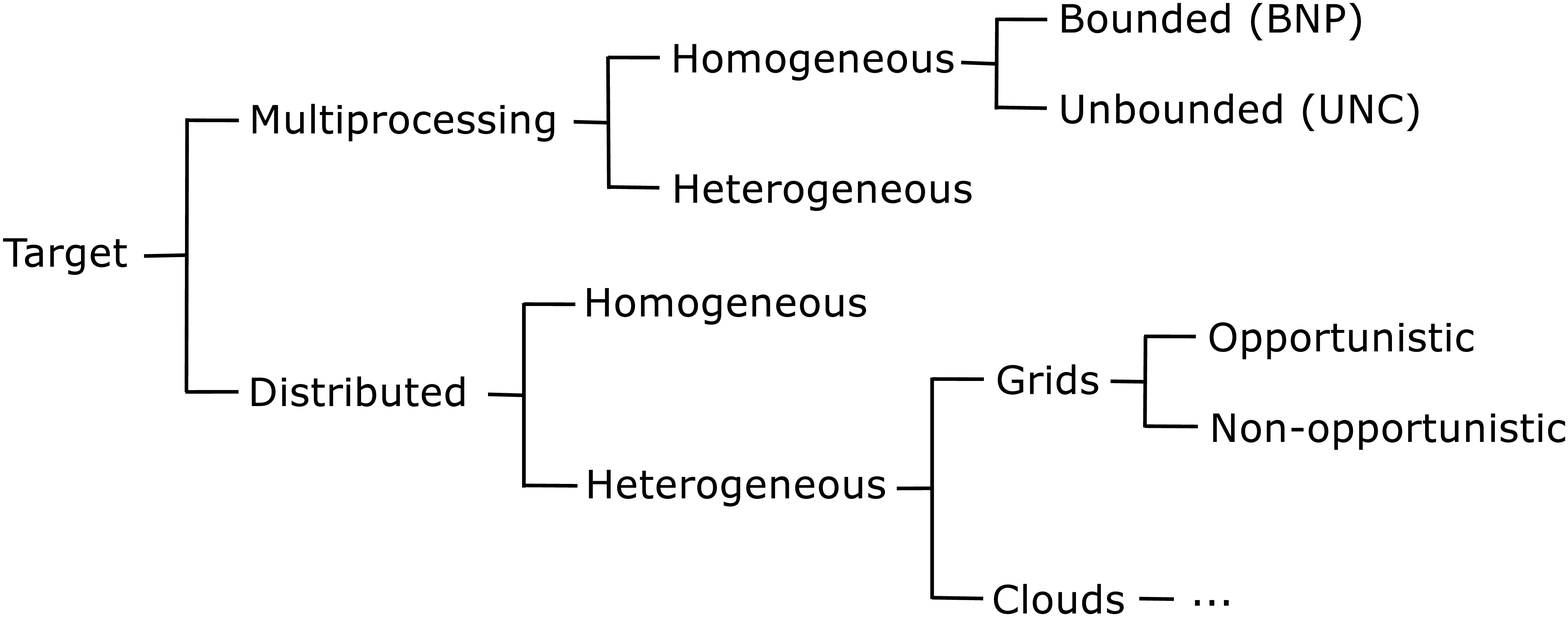}
	\caption{Taxonomy according to the target system.}
	\label{fig:tax_schedTarget}
\end{figure}

	The target system for a scheduler is associated to the computational architecture of the system where the scheduled jobs will run. Therefore, another factor that has influence on a scheduler design is information about the target system to make decisions. Two main characteristics considered during scheduling are the resources processing capacity and link bandwidth value interconnecting the processing resources. Our classification regarding the target system is shown in Figure~\ref{fig:tax_schedTarget}. 

	Although multiprocessing system may be apart from the distributed systems definition since they are shared memory systems, the study of scheduling algorithms for multiprocessors contributes to distributed systems in two aspects: (i) algorithms for multiprocessing systems can be adapted to work in distributed systems architectures; and (ii) multiprocessing systems may be part of distributed systems such as clusters, grids, or clouds. Scheduling targeting multiprocessing systems can be split in \emph{homogeneous multiprocessing} or \emph{heterogeneous multiprocessing}. Studies in unbounded homogeneous multiprocessing systems are common, such as the analysis made by Papadimitriou and Yannakakis in~\cite{Papadimitriou1988}. The so called unbounded schedulers assume that the number of processors is always greater than the number of tasks being scheduled, and this kind of algorithm had previously been classified as \emph{unbounded number of clusters} (UNC)~\cite{KwokStatic1999}. Conversely, scheduling algorithms considering bounded homogeneous multiprocessors, previously classified as \emph{bounded number of processors} (BNP)~\cite{KwokStatic1999}, do not make such assumption. Examples include the \emph{Modified Critical Path} (MCP) and the \emph{Earliest Task First} (ETF). Heterogeneous multiprocessing systems were less explored in the past than its homogeneous counterpart. 
	
	Considering the scheduling in distributed systems, i.e., with no shared memory, our classification comprises both \emph{homogeneous} and \emph{heterogeneous} systems. Works in homogeneous distributed systems generally consider them as computing clusters with identical machines interconnected by a homogeneous network. Chang and Oldham present models for dynamic task allocation in distributed homogeneous systems in~\cite{Chang1995}. A study on scheduling of parallel jobs in homogeneous distributed systems is presented by Karatza and Hilzes in~\cite{Karatza03}. Hagras and Jane\v{c}ek present an algorithm \emph{called critical nodes parent trees} (CNPT)~\cite{Hagras2003}. Large homogeneous datacenters gave momentum to scheduling in homogeneous systems~\cite{Tang2008}, including compositions of systems such as in multicluster grids~\cite{He:2004}.	
	
	Scheduling in heterogeneous distributed systems received substantial attention after the popularization of grid computing and, more recently, the cloud computing paradigms. Buyya et al. present the Nimrod-G~\cite{NimrodG}, which is a system for resource management and scheduling for grids. In addition, various simulators were developed to aid algorithms evaluation. Among the most cited ones are GridSim~\cite{GridSim}, developed by Buyya and Manzur, and the SimGrid~\cite{casanova2001simgrid}, developed by Casanova.
	
	Grid computing originally started as an opportunistic computing community-based platform for parallel processing. Heymann et al. proposed an algorithm for adaptive scheduling of master-worker applications in opportunistic grids~\cite{Heymann2000}. Another work that considers opportunistic grids is presented by Dutot in~\cite{Dutot05schedulingmoldable}. An architecture for scheduling in opportunistic grids was developed by Netto et al.~\cite{netto:rt-mac-200601}. 
	
	Grids evolved from desktop-based cycle-stealing mechanisms to more powerful processing platforms composed of shared heterogeneous beowulfs and clusters. Concerning those non-opportunistic grids, Boeres et al.~\cite{Boeres:2003} study the integration of static and dynamic algorithms for heterogeneous distributed systems, presenting experiments using hybrid (or semi-static) algorithms. Casanova et al. presented heuristics for the scheduling of parameter sweep applications in grids~\cite{heurGrads}. He et al. described a min-min heuristic for task scheduling in grids~\cite{He2003}.
	
	After the maturing of the computational grid, the cloud computing emerged as a new computing paradigm in distributed systems. The cloud branches in the taxonomy are discussed in Section~\ref{sec:clouds}.

\subsection{Application}
\begin{figure}[!htbp]
	\centering
	 \includegraphics[width=0.6\textwidth]{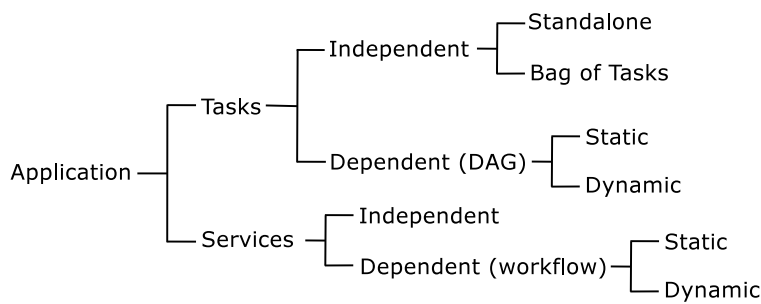}
	\caption{Taxonomy according to the application model.}
	\label{fig:tax_schedApp}
\end{figure}

	Application characteristics, as the target system, also have major influence in the development of scheduling algorithms. Therefore, it is worthwhile classifying the application models considered in the scheduling literature. Our proposed scheduling classification scheme according to the application to be scheduled is shown in Figure~\ref{fig:tax_schedApp}. In the most general classification, we separate the application model in \emph{tasks} or \emph{services}. A \emph{task} is a self-contained job that carries its own executable code. A task may also carry the necessary input data for its execution (or a reference to its location). A \emph{service} is a job which is actually an invocation to executable code that is available through an interface on the computational resource~\cite{FosterDynamic}. Therefore, a service call carries information about parameters and may also carry information on input data.
	
	 Both tasks and services can be classified as \emph{dependent} or \emph{independent}. Independent tasks or services perform no communication, while the dependent ones present communication dependencies among them. \emph{Independent tasks} can be either \emph{standalone} or \emph{bag of tasks} (BoT). While a standalone task has no relation to other tasks arriving to the scheduler, as the ones studied in~\cite{Maheswaran99}, tasks within a bag of tasks have similarities among them. An example of a class of BoT are the parameter sweep applications, where the same task is executed many times with different input parameters. Casanova et al. present algorithms for the scheduling of parameter sweep applications in~\cite{heurGrads}. Dependent tasks can be modeled as directed acyclic graphs (DAGs), in which nodes represent tasks and arcs represent communication dependencies among tasks. In our classification we separate DAGs into two classes: \emph{static DAGs} and \emph{dynamic DAGs}. Static DAGs have all their tasks run during their execution, such the ones scheduled by the algorithm presented by Zhao and Sakellariou~\cite{ZhaoHybrid}. On the other hand, dynamic DAGs may change during the execution: conditional tasks may not be run or new tasks may be added to the graph as a result of iterations in the DAG definition~\cite{Cicerre:2004}. Dynamic DAGs can also be modeled as Petri nets~\cite{YuTaxonomyJGC2005}. 
	 
	 The \emph{service oriented computing} (SOC) paradigm has expanded more recently. The same classes definition applied for tasks can be applied for services. However, we do not separate services in \emph{standalone} or \emph{bag of services}, since those terms are not used when dealing with service scheduling. When regarding to dependent services, which are also commonly modeled as DAGs, we call them \emph{workflows}. This term is now widely used when referring a set of dependent services and scientific applications to be scheduled in grids or clouds, as well as it is more ample and may include cycles.  

\subsection{Objective Function Optimization}
\label{sec:objectives}
	
	A scheduler can optimize different objectives when scheduling applications and their jobs in distributed systems. The minimization of the makespan is the most common optimization objective found in the scheduling literature \cite{PinedoScheduling}. In \cite{GridScheduling}, Dong and Akl classify the objective functions in either \emph{resource centric} or \emph{application centric}. Although this classification indeed reflects some aspects of the scheduling objectives, we prefer to approach this taxonomy in a different way because some objective functions cannot be said either resource centric or application centric. For example, minimizing data movement \cite{WorkflowDataMovement} can be thought as centered on both the resources pool and the application itself.
	
\begin{figure}[!htbp]
	\centering
	 \includegraphics[width=0.45\textwidth]{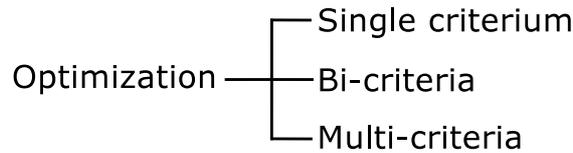}
	\caption{Taxonomy according to the optimization of the objective function.}
	\label{fig:tax_schedObjectiveFunction}
\end{figure}

	We separate the scheduler objective function optimization in distributed systems into three branches: single criterium, bi-criteria, and multi-criteria, as shown in Figure \ref{fig:tax_schedObjectiveFunction}.

	Single criterium scheduling algorithms try to find a schedule that optimizes one criterium. In this class of scheduling algorithms, it is possible to identify the most common objectives: minimize execution time \cite{HEFT}, minimize communication \cite{WorkflowDataMovement}, maximize throughput \cite{Bar-Noy:2002:ATM:586839.586857}, maximize load balancing \cite{Shirazi:1995:SLB:583069}, and maximize utilization \cite{MaxUtilization}. Bi- and multi-criteria algorithms focus on finding the best schedule following two or more (often conflicting) objectives. Optimizing more than one objective often leads to tradeoffs with no clear distinction among possible solutions to reach a single optimal solution. With no solution that is better than all the others, the algorithm must actually generate a set of solutions that are \textit{non-dominated}, meaning that for any solution in this set, there is no other solution that can improve one objective without worsening another. The \textit{Pareto front} is a set composed of non-dominated solutions.
	
	It is common to trade-off between two objectives in scheduling, yielding bi-criteria scheduling algorithms that optimize, for example, reliability and execution time~\cite{DoganBiobjective}, workflow bi-criteria optimization~\cite{WieczorekBicriteria}, robustness and makespan~\cite{CanonBicriteria}, service instantiation costs and makespan~\cite{BittencourtBicriteria}, execution time and costs~\cite{Bessai2012}, and resources utilization and makespan~\cite{BittencourtMGC2012}. Multi-criteria scheduling also frequently involves optimizing execution time along with a set of other criteria~\cite{WieczorekMulticriteriaTaxonomy2008}.

\paragraph{Optimization techniques} 
	Classification and evaluation of scheduling techniques in distributed systems have been addressed before in the literature \cite{CasavantTaxonomy1988,KwokStatic1999, HamscherEvaluation2000}. The na\"ive approach to perform the scheduling is to try all possible combinations and return the best one. This is feasible only if the size of the input instance (e.g. amount of tasks and/or dependencies and amount of computational resources) is sufficiently small, since there exist an exponential number of combinations and, consequently, large instances are unfeasible within a reasonable wall time. That is why scheduling is often approached by using sub-optimal techniques, such as heuristics~\cite{CanonComparative} and meta-heuristics~\cite{Pandey2010}, as well as combinatorial optimization methods~\cite{Genez2012} and approximation algorithms~\cite{6847985}.
	
\subsection{Output}
\begin{figure}[!htbp]
	\centering
	 \includegraphics[width=0.27\textwidth]{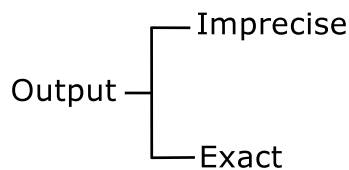}
	\caption{Taxonomy according to the scheduling output.}
	\label{fig:tax_schedPrecision}
\end{figure}

	A scheduling algorithm produces a schedule can be either is \emph{exact} or \emph{imprecise}, depending on the expected input. When the produced schedule is  \emph{exact}, it is a consequence of an \emph{exact} input data, thus producing a final schedule that will be fully obeyed. On the other hand, if the schedule produced is assumed to be \emph{imprecise}, the algorithm should make assumptions to deal with stochastic and uncertain input data, which includes application demands and resources availability imprecisely specified or estimated. The scheduling output taxonomy is shown in Figure~\ref{fig:tax_schedPrecision}.

	Many scheduling algorithms consider that the produced schedule is exact, as a result of the input data to be precise~\cite{KwokStatic1999, HEFT, heurGrads}. Some algorithms assume that other mechanisms are responsible for improving the scheduling quality if the input data is not accurate (or imprecise). Such mechanisms include reactive rescheduling, as in dynamic or semi-static scheduling approaches. Note that the imprecisions can result from user specification or performance measurement/prediction systems, such as in~\cite{Gomes2011}.

\section{Scheduling in Cloud Computing}
\label{sec:clouds}
	Hitherto we described characteristics of the scheduling problem and target distributed systems in general. Although the presented overview is broad, the concepts are directly applicable to specific computer systems. Henceforth, we specialize this broader view to the cloud computing paradigm. In this section, we present how the scheduling problem in cloud computing has been developed, and discuss scheduling peculiarities and major challenges in cloud computing. Moreover, we extend some aspects of the proposed taxonomy to incorporate cloud computing characteristics and present how to tackle with schedulers objectives in clouds.
	
	Scheduling algorithms and their associate performance are strongly influenced by the input data and specific characteristics of the target system. As different service models implemented in cloud computing have specific objectives, the design of new scheduling algorithms is a clear need to produce efficient services. Actually, it is common that conflicting objectives can be part of a single optimization goal.

\subsection{Cloud computing service models}
	In the cloud computing model, users are not aware of the specific technology employed to enable service provision. Such abstraction results in the provision of services over the Internet which are dynamic and scalable~\cite{NIST}. The cloud computing business model allows clients to extend their computing platforms by leasing virtualized computational resources offered as services by cloud computing providers through the Internet. This resource leasing provides elasticity to the client computational power by running software or development platforms, or by leasing virtualized hardware. Cloud computing has three canonical models: software as a service (SaaS), platform as a service (PaaS), and infrastructure as a service (IaaS). Other service models encountered in the literature are often specialization or combination  of these three models.

\begin{figure}[!htbp]
	\centering
	 \includegraphics[width=0.45\textwidth]{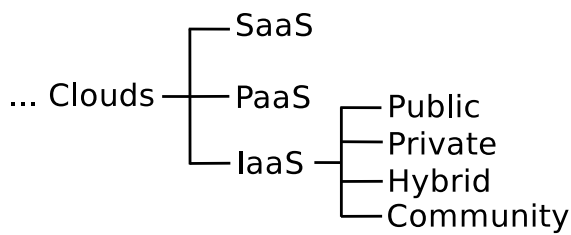}
	\caption{Cloud extension of the taxonomy according to the target system.}
	\label{fig:tax_schedTargetCloud}
\end{figure}

	The IaaS is a popular model from a resource management perspective, thus scheduling in such systems has received great attention in the research community~\cite{denBossche2010, 6782394, bittencourt2012scheduling}. In a nutshell, IaaS is a cloud computing model where virtual servers are made accessible to clients through the Internet. In terms of access policy, an IaaS can be \emph{public}, \emph{private}, or \emph{hybrid}. A \emph{public IaaS} offers servers in a pay-per-use basis, while a \emph{private IaaS} can be seen as a virtualized and more user-transparent cluster. As a combination of these two, a \emph{hybrid IaaS} comprises both public and private IaaS clouds, offering private resources that can be expanded with paid public cloud resources, realizing the so-called \emph{elasticity}~\cite{BittencourtJISA}. Another model considered in the literature is the \textit{Community Cloud}, being a composition of private clouds sharing their resources through their administrative borders. To cover these types of clouds, an extension of the target system taxonomy is shown in Figure \ref{fig:tax_schedTargetCloud}. 
	
	Clearly, the classification of cloud access in public, private, hybrid, or community can be extended for PaaS and SaaS. In special, community clouds inherit from grid computing many challenges related to administrative aspects of access control that are discussed in a multi-cloud framework~\cite{calheiros2012aneka}. In our classification, we put these types of cloud access explicitly under IaaS since it is at the infrastructure level that the resource sharing effectively occurs.
	
\subsection{Client and provider perspectives}
	Cloud management,  which includes scheduling, must deal with two different perspectives: one from the cloud client and the other from de cloud provider. Cloud clients demand quality of service and low prices, while cloud providers focus on offering QoS and making profit. The development of algorithms for each perspective requires different approaches to achieve such different optimization objectives.
		
\begin{figure*}[!htbp]
\centering
\subfigure[Client perspective.]
{
 \includegraphics[width=0.47\textwidth]{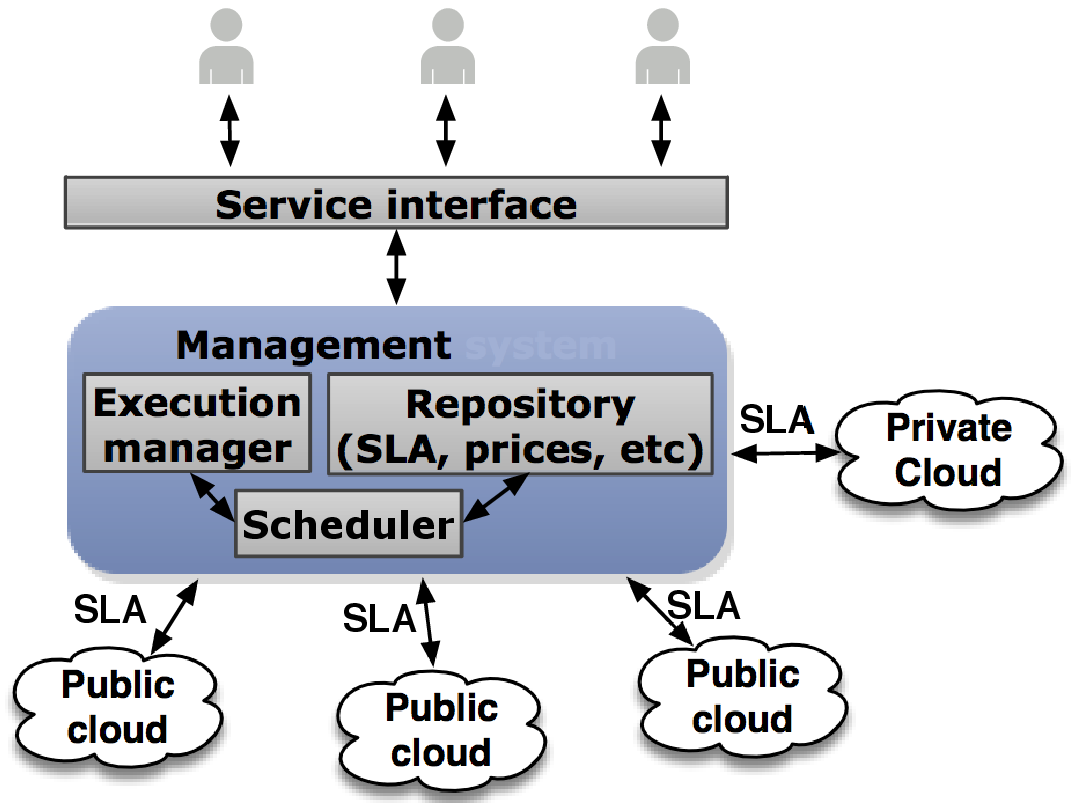}
 \label{fig:userArc}
 }
\subfigure[Provider perspective.]
{
 \includegraphics[width=0.47\textwidth]{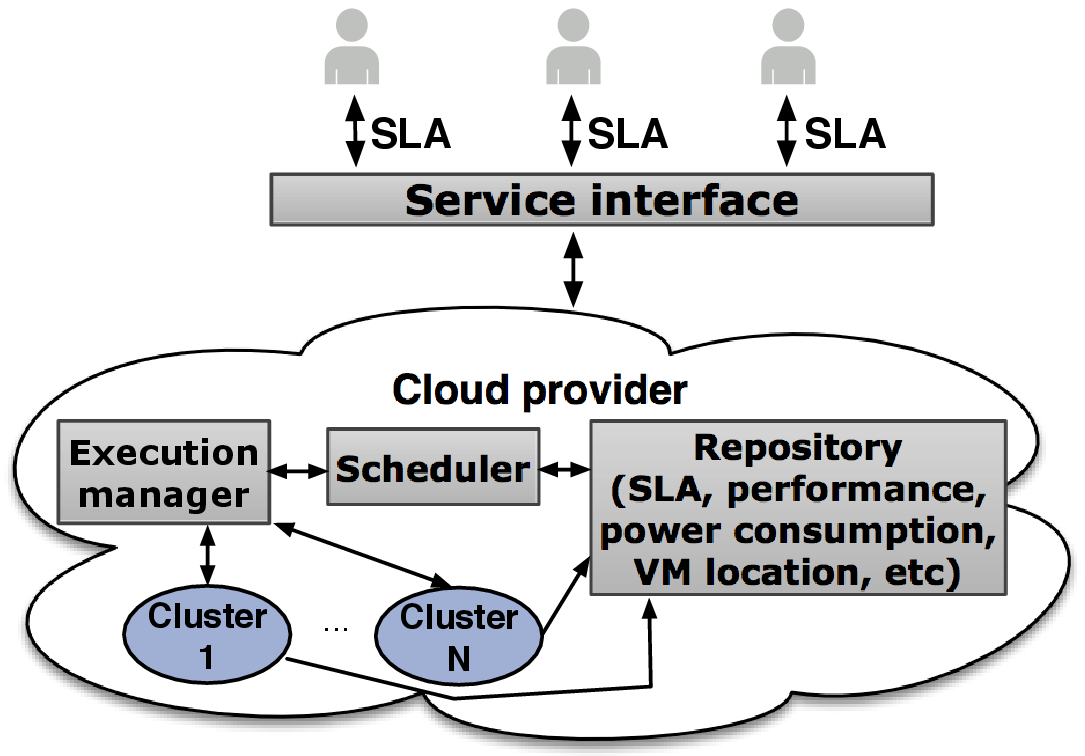}
 \label{fig:providerArc}
}
\caption{Client and provider perspectives in cloud computing.}
\end{figure*}

	Figure \ref{fig:userArc} illustrates the client perspective, in which users have access to a private or in-organization task/service submission interface. The management system is responsible for gathering and storing information about available clouds, as well as their services and prices. In this perspective, clients can utilize services from a private cloud (if available) and/or from multiple public clouds,  which is consistent with the taxonomy shown in Figure \ref{fig:tax_schedTargetCloud}. In what concerns the scheduler, it must deal with the trade-off between quality of service and price, which depends on users requirements and objectives.

	The provider perspective is illustrated in Figure \ref{fig:providerArc}. The service interface can be located in the cloud provider or in a client broker, usually offered as a portal, a set of tools or as an API (application programming interface). Based on the client/application requirements defined in the SLAs, the cloud provider must decide the best resource configurations to be used according to its profitability principles\footnote{As a market-oriented business, the provider can focus on different strategies, from strong confidence relations with its clients to lower level services. The discussion on business/market strategies is very complex and is out of the scope of this paper.}. Depending on the service level offered by the provider interface (IaaS, PaaS, SaaS), the scheduler will cope with different inputs for the decision-making. In the case of an IaaS provider, the scheduler is responsible for allocating virtual machine requests on the physical substrate. In the case of SaaS or PaaS provider, the scheduler allocates client applications and/or application requests to the virtualized computing resources.

\begin{figure}[!htbp]
 \centering
 \includegraphics[width=0.6\textwidth]{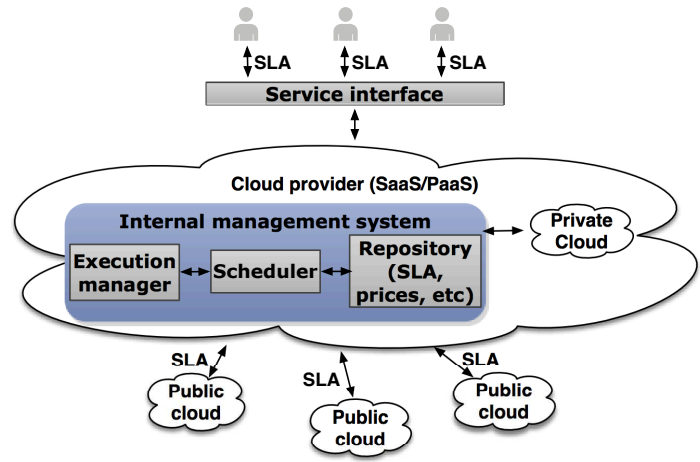}
 \caption{Cloud management with 2 (or more) SLA layers.}
 \label{fig:2levelsArc}
\end{figure}

\begin{figure}[!htbp]
	\centering
	 \includegraphics[width=0.45\textwidth]{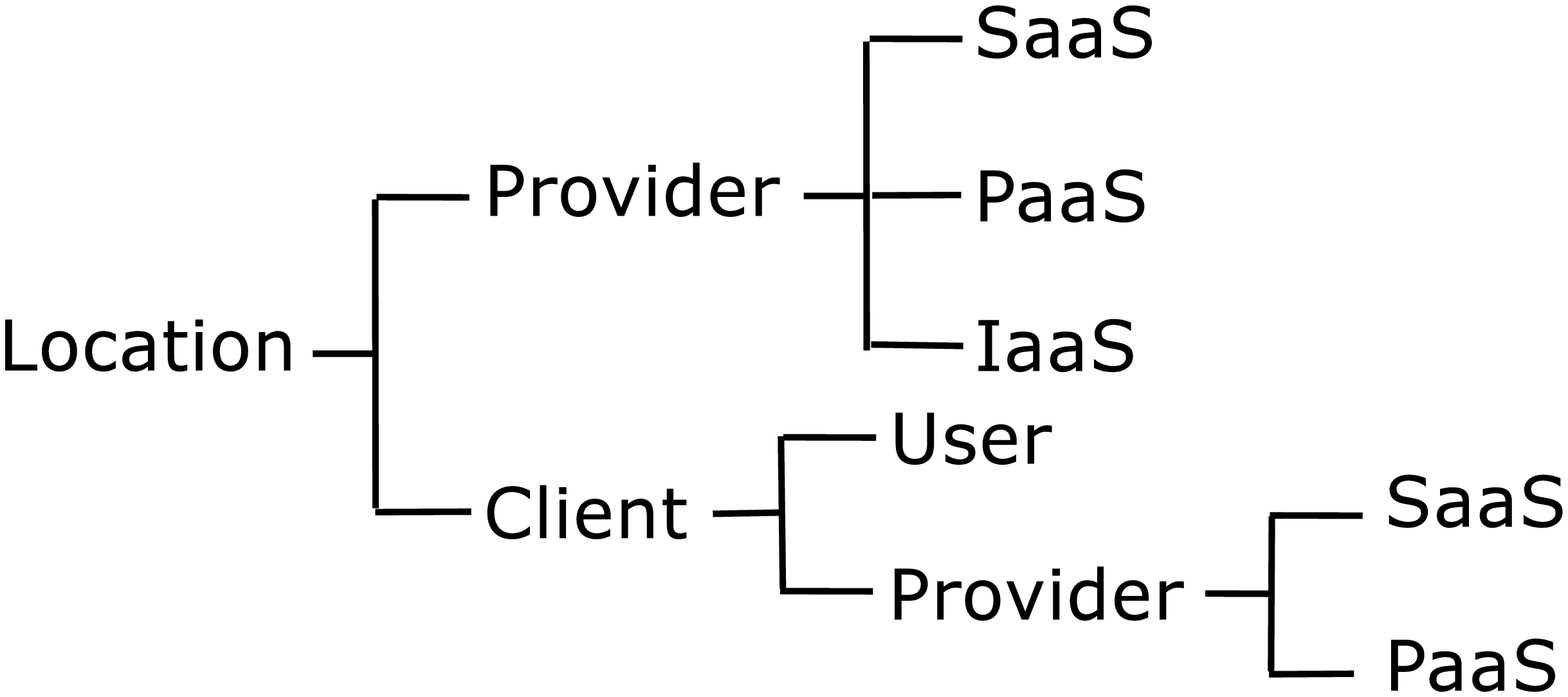}
	\caption{Taxonomy in clouds according to the scheduler location.}
	\label{fig:tax_schedCloudTarget}
\end{figure}

	A third perspective, illustrated in Figure \ref{fig:2levelsArc}, combines both client and provider perspectives, in which a public cloud provider makes use of other public clouds to provide services to its clients. In this perspective, two SLA layers exist: one between the provider and its clients, and one between the provider and the other cloud providers. In this scenario, different terms can exist at each SLA level, which must be translated from one layer into QoS requirements at the other layer so the scheduler can allocate client requests. Therefore, both client and provider perspectives co-exist, and the scheduler must make decisions on how to get the best QoS according to the current demand, but also should focus on profitability when acting as a client from other public clouds. Figure \ref{fig:tax_schedCloudTarget} presents a taxonomy for the scheduler location in the cloud. 
	
\subsection{Scheduler ``Feedstock''}
\begin{figure}[!htbp]
\subfigure[Application model.]
{
	 \includegraphics[width=0.5\textwidth]{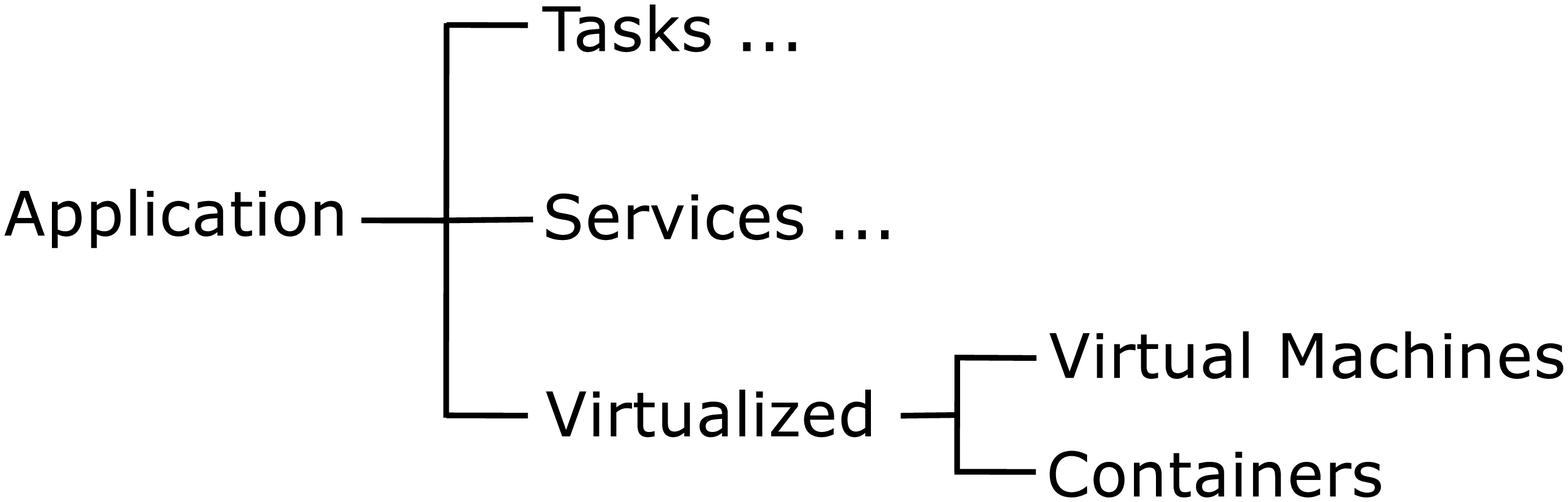}
	\label{fig:tax_schedAppCloud}
}
\subfigure[Input semantics.]
{
	 \includegraphics[width=0.5\textwidth]{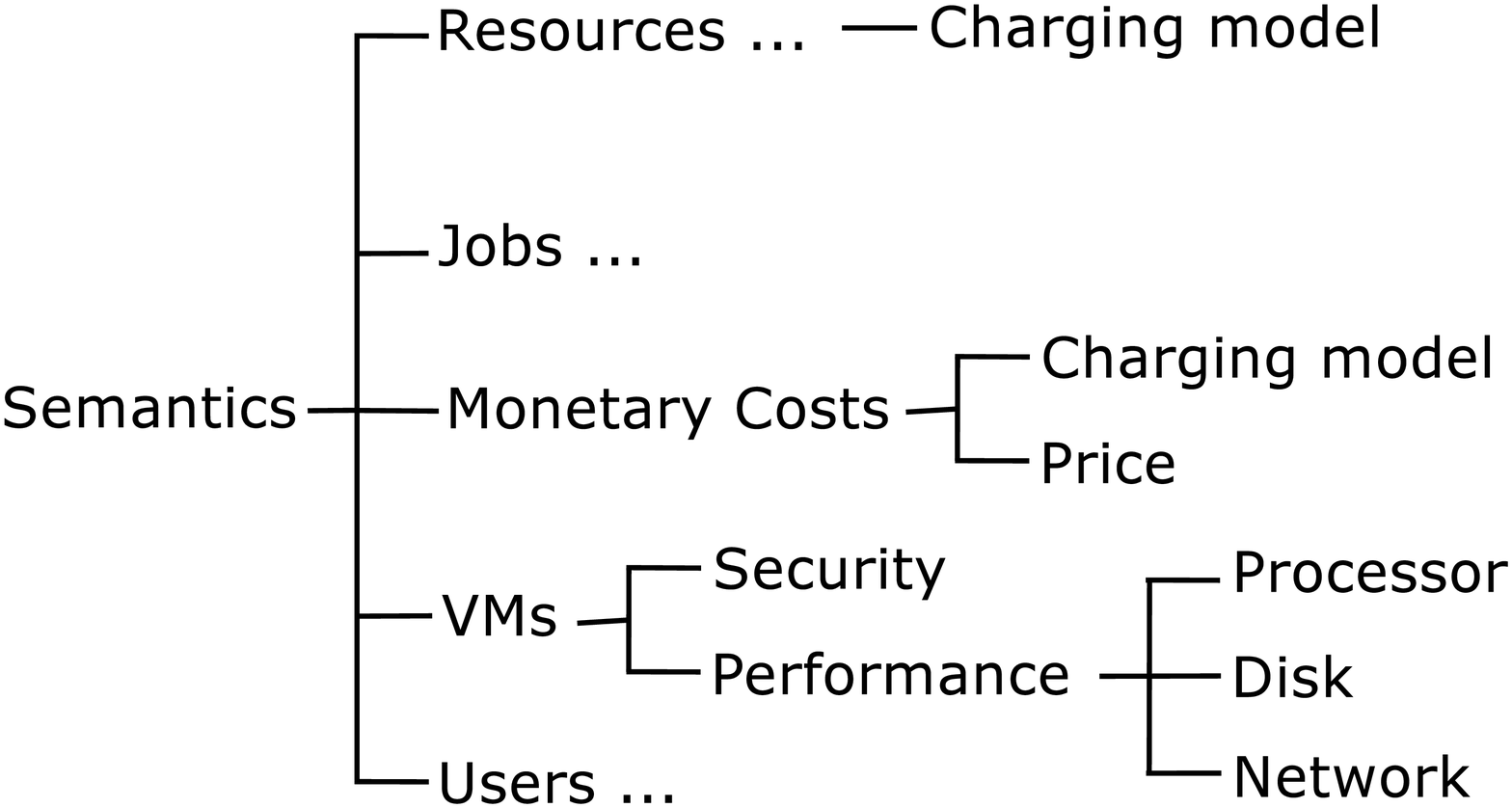}
	\label{fig:tax_InputSemanticsCloud}
}
\caption{Application model and input semantics taxonomy extensions for cloud computing.}
\end{figure}

	A fundamental difference between clouds and previous distributed systems is the wide adoption of virtualization to offer computing resources to the users. To achieve efficient server consolidation, VM allocation (or placement) must be performed, i.e., a scheduling decision that is different from the application scheduling. In this section, we denominate \emph{scheduler feedstock} as the type of computer software being considered by the scheduler: virtual machines or applications. Thus, when developing schedulers for cloud computing, we must specify if the scheduler is a \emph{VM scheduler} or an \emph{application scheduler}. These two different resource allocation facets in clouds have implications that were not present in grids and clusters. Based on this, we derive a new branch in the the application model taxonomy in Figure \ref{fig:tax_schedAppCloud} (previously shown in Figure \ref{fig:tax_schedApp}).  Moreover, we extend the input semantics taxonomy (previously shown in Figure \ref{fig:tax_InputSemantics}) to cover input data about virtualized environments. The new branch of the input semantics is shown in Figure \ref{fig:tax_InputSemanticsCloud}.

	The VM scheduler needs information on the resource requirements of the VMs being allocated, namely security constraints and hardware requirements (CPU performance and number of cores, amount of RAM, etc.). Another relevant information for the scheduler is the charging model of the VMs. The charging model have implications in the expected duration of the allocation and also in its preemption rules~\cite{Vieira:2013aa}. For example, Amazon EC2 spot instances can be interrupted by the provider without prior notice, while doing that for on-demand instances would mean an SLA violation from the provider. Taking this information into account can help the VM scheduler to properly place VMs onto the physical machines.

	To deal with the users requests, an application scheduler is necessary when the scheduler is located in the cloud client, as shown in Figure \ref{fig:userArc}.  This scheduler decides on which VMs the applications will run. If the VMs are already deployed and running, the application can be sent for execution. Otherwise, the private cloud and/or the public clouds that will be used to run the application must deploy new virtual machines according to the application scheduler output. Thus, the virtual machine scheduler present in the cloud (Figure \ref{fig:providerArc}) must be invoked to decide on which physical machine the new VM(s) will be placed. After that, all VMs can be switched on and then the application components can run.
	
	When considering IaaS providers, one important aspect that is made clear from the two scenarios in Figures \ref{fig:userArc} and \ref{fig:providerArc} is that application and VM schedulers information exchange is limited. The application scheduler can ask for a certain amount of VMs with a set of characteristics to the VM scheduler, and no further information is exchanged. In other words, the VM scheduler has no specific information about the application(s) that will run on those VMs. As a consequence, the resulting VM scheduling in the physical resources can, in practice, lead to low resource utilization even if all physical resources in a machine are currently allocated to virtual machines. However, if we consider, for instance, a two or more SLA layers scenario with an SaaS or PaaS provider (Figure \ref{fig:2levelsArc}), the application scheduler and the private cloud VM scheduler are under the same administrative domain. Therefore, information about the computing demands of the application being scheduled can serve as input to the VM scheduler to improve allocation. By considering the application demands and the VM requirements altogether, the VM scheduler can, for example, allocate a VM that have low CPU utilization and high I/O demands in the same physical machine of a VM with high CPU utilization and low I/O demands, avoiding bottlenecks that could impair quality of service, as well as maximizing hardware utilization.

\subsection{Scheduler objectives}
\label{sec:objectivesCloud}
\begin{figure}[!htbp]
	\centering
	 \includegraphics[width=0.65\textwidth]{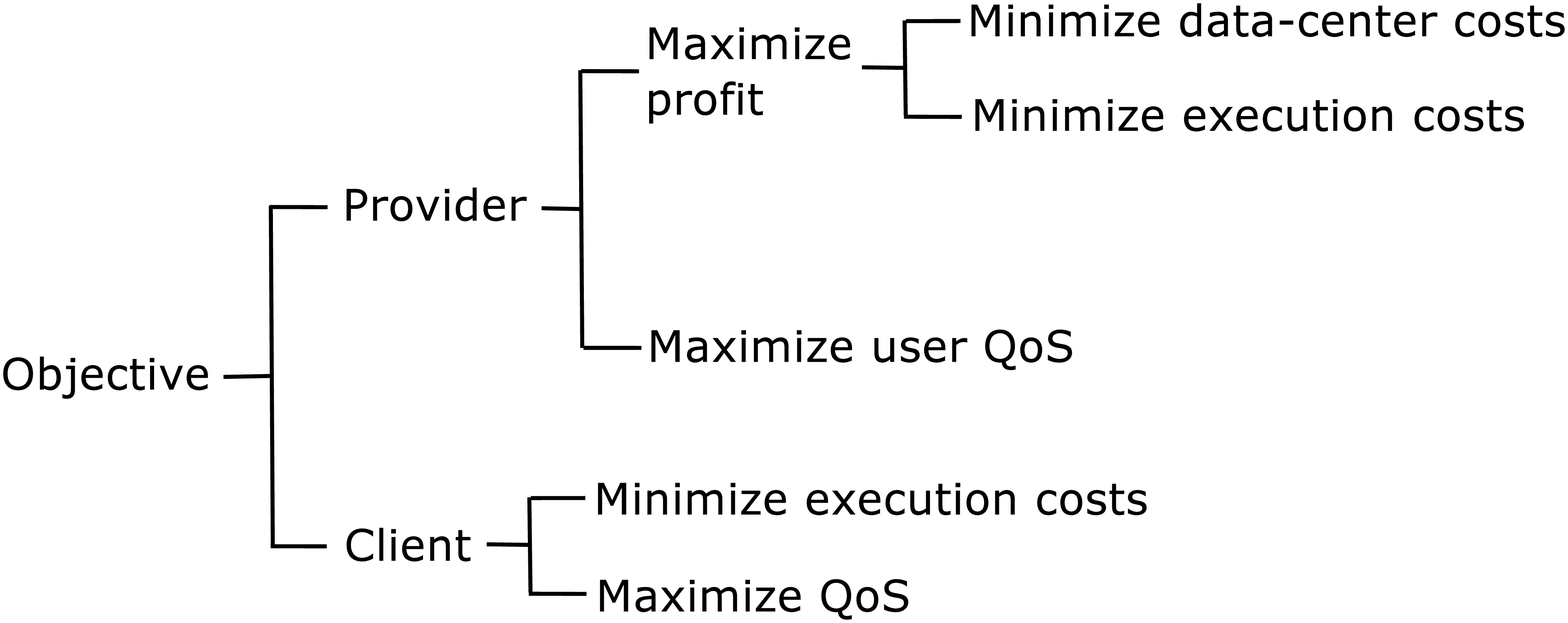}
	\caption{Taxonomy in clouds according to the scheduler objectives.}
	\label{fig:tax_schedCloudObjectives}
\end{figure}

	In clouds, the objectives pursued by a scheduling algorithm often depend on where the scheduler is running. Different approaches and algorithms must be implemented to achieve these objectives in a variety of scenarios. Each scheduling approach, according to the scheduler location, takes as input information of diverse granularity and semantics. In order to reach an objective, the scheduler must optimize one or more criteria following an objective function (see Figure \ref{fig:tax_schedObjectiveFunction}). Thus, a single objective can be pursued by an algorithm which optimizes one criterium, but can be one among multiple objectives to optimize (multiple criteria).  The proposed taxonomy for scheduler objectives in clouds is presented in Figure \ref{fig:tax_schedCloudObjectives}. Its details are discussed in the next sections.

\subsubsection{Providers}
	Cloud providers must supply its clients with virtualized computing resources according to a service level agreement. In the case of IaaS providers, it receives VM requests from its clients and must schedule them on the underlying physical machines following the resource requirements of these VMs, such as amount of processing cores, performance, disk space and speed, networking requirements, and so on. The virtual machine placement decided by the scheduler will have direct impact in the datacenter utilization and VM performance observed by the clients. Therefore, a scheduler at this level is concerned with minimizing datacenter costs subject to VM specifications in the SLAs established with the clients. From the input semantics for clouds (Figure \ref{fig:tax_InputSemanticsCloud}), we observe that this SLA can include security and hardware requirements, characteristics of the charging model to be followed, and application demands.

	To \textit{minimize datacenter costs}, cloud providers can rely on algorithms that, for instance, minimize the amount of active machines. This can be achieved by scheduling VMs on the minimum possible amount of physical machines as soon as the desired (or ensured by SLA contracts) quality of service is maintained. However, as the clients want elasticity, the number of running VMs changes over time, increasing with new VM requests and decreasing when clients turn their VMs off. This leads to allocations where hardware is underutilized, whereas virtual machine migration can remediate this situation. Thus, the scheduling frequency (Figure \ref{fig:tax_schedFrequency}) of VMs in the datacenter must consider how volatile the VM requests are. For instance, a periodic rescheduling or event-driven rescheduling can be utilized to maintain a high hardware utilization.

	One way of  minimizing running costs is through profiling running virtual machines on-the-fly, aiming at discovering the VMs resource usage patterns. This can lead to a better reallocation of virtual machines, avoiding to collocate virtual machines that will potentially create bottlenecks in the underlying hardware. For example, two I/O intensive virtual machines could be placed in different physical hardware, while two VMs that have strong communication relations could be placed in the same physical machine. The VM characterization problem is yet a challenge that can be addressed through hardware and software monitoring in order to find use patterns, and the development of new algorithms should consider such patterns to avoid overload is desirable. However, privacy and security concerns must be taken into consideration when monitoring VMs.

	We identified two main objectives from schedulers located in cloud providers: maximize profit and maximize quality of service. Profit maximization can be achieved by adopting more specific scheduling objectives, such as minimizing the datacenter running costs and minimizing the tasks execution costs. Consolidation can be utilized for both objectives: cloud providers can adopt energy-aware resource allocation policies to minimize datacenter running costs, or they can follow a consolidation objective function that focuses on minimizing the amount of cloud resources (rented from other providers as well) necessary to run a set of tasks. 
	
	Energy consumption has attracted attention from the resource allocation community in cloud data centres~\cite{Ghribi2013,Wu2014141,Wang201491,TchernykhOnline2014} as part of the so-called \textit{Green Computing} concept~\cite{GreenComputing}. Energy-aware resource allocation commonly takes the form of a VM placement optimization in IaaS data centers, where the VM allocation is determined by quality of service requirements (usually defined in an SLA with the cloud client) and consolidation of VMs into the physical machines available. Therefore, consolidation helps in reducing the amount of machines powered on, and thus resource allocation collaborates to reduce power consumption. Energy-aware allocation can also take into account power consumption of network devices, aiming at placing VMs closer to data sources to reduce network traffic~\cite{Wang201491}.
	
	Clouds offering platform (PaaS) and software (SaaS) as services need a computational infrastructure to host the platform and run the software services. These cloud providers can own the whole infrastructure or they can rely on IaaS providers to achieve elasticity, shrinking their infrastructure and reducing costs whenever possible. In this scenario, the SaaS or PaaS provider composes a hybrid cloud to run services offered to its clients, therefore resulting in two SLA layers. As in other businesses, different strategies can be used by the SaaS/PaaS provider to attract clients, such as improved service quality with higher charges or cheap services with lower service quality. The scheduling algorithm must work with different optimization functions depending on the desired strategy. For example, if the cloud provider wants to offer a better service, the scheduler may focus on the maximization of pre-established quality criteria of the underlying IaaS providers. On the other hand, the cloud provider can focus on reducing its running costs, where a cost-oriented scheduler could be used to minimize the budget when renting IaaS resources according to client SLAs. As a consequence, PaaS and SaaS providers can follow the objectives \textit{minimize execution costs} and \textit{maximize user QoS}, as shown in Figure \ref{fig:tax_schedCloudObjectives}.
	
	Clearly, some of the objectives described here can overlap with each other. How and when to combine such objectives, as well as algorithms for such combinations, are still challenges to be addressed. An even more challenging objective to be considered by a scheduler is the minimization of maintenance and personnel costs, which involves man-hour costs and expected lifetime of datacenter components.  
	
\subsubsection{Clients}
	The final cloud client can be a user that utilizes exclusively either private or public cloud resources, or a user that composes a hybrid cloud to run his/her applications. In both cases, the scheduler objective should be to reduce costs while receiving a satisfactory QoS according to the user/applications requirements. Client QoS in clouds is often measured in terms of response time, performance, and availability, but other requirements regarding resources configuration can also be part of the SLA contract. As in other market niches, scheduling from client perspective usually involves a trade-off between budget and quality (QoS parameters).
	
	From the client perspective, research has been focused in two main objectives: maximize QoS (often minimize the makespan) and minimize the monetary costs involved when using the public cloud. These objectives can be pursued separately, when only one of them matters for the scheduling, or together, when both objectives are related. One example is the minimization of the monetary costs but at the same time respecting a maximum makespan for the workload.
	
	When we consider the client as the owner of a private cloud, scheduling objectives can have similarities with the scheduling from the provider perspective in the sense that the number of active physical resources should be minimized, avoiding unnecessary monetary costs. However, scheduling in the private cloud is mainly focused on performance metrics instead of profitability. This leads to the development of multi-criteria approaches to deal with the scheduling in hybrid clouds, which needs to address both private cloud objectives and cost minimization to compose the hybrid cloud.

\section{Literature review on scheduling in clouds}
\label{sec:survey}
	This section presents a literature review of the scheduling problem in cloud computing. We followed a simple search protocol using online indexing systems\footnote{Google Scholar, IEEExplore, ACM Digital Library, Scopus}. Searches were performed for each year, from 2010 to 2016, using the filtering tools available in the indexing online systems. Variations of the terms \textit{cloud computing}  and \textit{scheduling} were appended to the proposed taxonomy branch names to find relevant work for the taxonomy branch problems. For example, to search papers on scheduling for SaaS, the search term \textit{scheduling cloud SaaS} was used. Results were collected and manually filtered to remove manuscripts from other areas (e.g. Physics/weather or works focused on e-Science applications, not in the scheduling problem). Most relevant papers were then selected having as criteria the focus of the text (should be focused on scheduling), venue relevance, and/or number of citations, resulting in a list of $18$ papers published in $2010$, $14$ in $2011$, $12$ in $2012$, $16$ in $2013$, $30$ in $2014$, and $11$ in $2015$, and $13$ in $2016$, totalling $114$ selected research papers we considered relevant to the scheduling problem in cloud computing systems.
	
	Selected papers were classified according to the taxonomy proposed in this paper, considering both the cloud (Section \ref{sec:clouds}) and pre-cloud (Section \ref{sec:pre-taxonomy}) taxonomy branches. This detailed classification is shown in Tables \ref{tab:surveyCloud1}, \ref{tab:surveyCloud2}, \ref{tab:surveyPreCloud1}, and \ref{tab:surveyPreCloud2}, where column titles are the short text notations introduced in the full taxonomy overview in the beginning of this paper (Figure \ref{fig:tax_full}). In these tables, a filled circle means the research paper listed in the row is relevant to the branch listed in that column. A filled circle does not necessarily mean the work in that row explicitly mentions the branch keyword, but states that the branch is somehow considered by that research. These tables aim at helping readers to find unexplored branches and combinations of topics that lack research, but also help in finding relevant related work that can be used as comparison and/or extensions to the readers research.

\begin{table*}[!htbp]
\thisfloatpagestyle{empty}
\fontsize{0.18cm}{0.16cm}\selectfont
\caption{Reviewed research articles from 2014 to 2016 classified considering the \textbf{cloud branches} of the proposed taxonomy.}
\label{tab:surveyCloud1}
\begin{center}
\setlength\tabcolsep{4pt}
\begin{tabular}{l|cccccc|ccccc|cc|cc|cccc|cccccc}
    \toprule
        & \multicolumn{25}{c}{Cloud taxonomy branches} \\[5pt]
        & \rot{S/T/.../C/SaaS} & \rot{S/T/.../C/PaaS} & \rot{S/T/.../C/IaaS/Public} & \rot{S/T/.../C/IaaS/Private} &  \rot{S/T/.../C/IaaS/Hybrid} & \rot{S/T/.../C/IaaS/Community}
        & \rot{S/OBJ/P/MP/Min. DC costs} & \rot{S/OBJ/P/MP/Min. ex. costs} & \rot{S/OBJ/P/Max. user QoS} & \rot{S/OBJ/C/Min. ex. costs} & \rot{S/OBJ/C/Max. QoS} 
        & \rot{S/I/S/MC/Charging Model} & \rot{S/I/S/MC/Price} &  \rot{S/A/V/Virtual machines}  & \rot{S/A/V/Containers} &  \rot{S/I/S/VMs/Security}  & \rot{S/I/S/VMs/P/Processor} & \rot{S/I/S/VMs/P/Disk} & \rot{S/I/S/VMs/P/Network} 
        & \rot{S/L/P/SaaS} & \rot{S/L/P/PaaS} & \rot{S/L/P/IaaS} & \rot{S/L/C/User} & \rot{S/L/C/P/SaaS} & \rot{S/L/C/P/PaaS--Broker} \\
        \midrule
\rowcolor{Grey7} \textbf{2016} & & & & & & & & & & & & & & & & & & & & & & & & & \\        
Abdullahi, M. et al. \cite{Abdullahi2016640}  & $\circ$ & $\circ$ & $\circ$ & $\bullet$ & $\circ$ & $\bullet$ & $\circ$ & $\circ$ & $\circ$ & $\circ$ & $\bullet$ & $\circ$ & $\circ$ & $\circ$ & $\circ$ & $\circ$ & $\bullet$ & $\circ$ & $\circ$ & $\circ $ & $\circ$ & $\circ$ & $\circ$ & $\circ$ & $\bullet$ \\
\rowcolor{Grey9} Bryk, P. et al. \cite{Bryk2016} & $\circ$ & $\circ$ & $\bullet$ & $\bullet$ & $\circ$ & $\circ$ & $\circ$ & $\circ$ & $\circ$ & $\bullet$ & $\bullet$ & $\circ$ & $\bullet$ & $\circ$ & $\circ$ & $\circ$ & $\bullet$ & $\bullet$ & $\bullet$ & $\circ $ & $\circ$ & $\circ$ & $\bullet$ & $\circ$ & $\circ$ \\
Deldari, A et al. \cite{Deldari2016}  &   $\circ$ & $\circ$ & $\circ$ & $\circ$ & $\bullet$ & $\circ$ & $\circ$ & $\circ$ & $\circ$ & $\bullet$ & $\bullet$ & $\circ$ & $\bullet$ & $\circ$ & $\circ$ & $\circ$ & $\bullet$ & $\circ$ & $\bullet$ & $\circ$ & $\circ$ & $\circ$ & $\bullet$ & $\circ$ & $\bullet$ \\
\bottomrule
\end{tabular}
\end{center}
\label{tab1}
\end{table*}%

\begin{table*}[!htbp]
\fontsize{0.18cm}{0.16cm}\selectfont
\caption{Reviewed research articles from 2014 to 2016 classified considering the \textbf{pre-cloud branches} of the proposed taxonomy.}
\label{tab:surveyCloud2}
\begin{center}
\setlength\tabcolsep{4pt}
    \begin{tabular}{l|ccc|ccc|ccc|cccc|cccc|ccc|cc}
    \toprule
        & \multicolumn{22}{c}{Pre-cloud taxonomy branches} \\[5pt]
        & \rot{S/O/Centralized} & \rot{S/O/Decentralized} & \rot{S/O/Hierarchical} & \rot{S/I/A/Uncertain} &  \rot{S/I/A/Stochastic} & \rot{S/I/A/Precise}
        & \rot{S/I/S/Resources} & \rot{S/I/S/Jobs} & \rot{S/I/S/Users} & \rot{S/F/Online} & \rot{S/F/Offline/Static} 
        & \rot{S/F/Offline/Semi-Static}  & \rot{S/F/Offline/Dynamic} & \rot{S/A/T/Independent} & \rot{S/A/T/Dependent} & \rot{S/A/S/Independent} & \rot{S/A/S/Dependent}
        & \rot{S/OPT/Single criterium} &  \rot{S/OPT/Bi-criteria} &  \rot{S/OPT/Multi-criteria} & \rot{S/OUT/Imprecise} & \rot{S/OUT/Exact} \\ 
         \midrule
\rowcolor{Grey7}\textbf{2016} & & & & & & & & & & & & & & & & & & & & & & \\
Abdullahi, M. et al. \cite{Abdullahi2016640}   & $\bullet$	& $\circ$	& $\circ$	& $\circ$	& $\circ$	& $\bullet$	& $\bullet$	& $\bullet$	& $\circ$	& $\circ$	& $\bullet$	& $\circ$	& $\circ$	& $\bullet$	& $\circ$	& $\circ$	& $\circ$	& $\bullet$	& $\circ$	& $\circ$	& $\circ$	& $\bullet$\\
\rowcolor{Grey9} Bryk, P. et al. \cite{Bryk2016}   & $\bullet$	& $\circ$	& $\circ$	& $\bullet$	& $\circ$	& $\circ$	& $\bullet$	& $\bullet$	& $\circ$	& $\circ$	& $\bullet$	& $\circ$	& $\circ$	& $\circ$	& $\bullet$	& $\circ$	& $\circ$	& $\circ$	& $\bullet$	& $\circ$	& $\circ$	& $\bullet$\\
Deldari, A et al. \cite{Deldari2016}  & $\bullet$	& $\circ$	& $\circ$	& $\circ$	& $\circ$	& $\bullet$	& $\bullet$	& $\bullet$	& $\circ$	& $\circ$	& $\bullet$	& $\circ$	& $\circ$	& $\circ$	& $\bullet$	& $\circ$	& $\circ$	& $\circ$	& $\bullet$	& $\circ$	& $\circ$	& $\bullet$\\
\rowcolor{Grey9} Gupta, I. et al. \cite{Gupta2016}    & $\bullet$	& $\circ$	& $\circ$	& $\circ$	& $\circ$	& $\bullet$	& $\bullet$	& $\bullet$	& $\circ$	& $\circ$	& $\bullet$	& $\circ$	& $\circ$	& $\circ$	& $\bullet$	& $\circ$	& $\circ$	& $\circ$	& $\bullet$	& $\circ$	& $\circ$	& $\bullet$\\
\bottomrule
\end{tabular}
\end{center}
\label{tab2}
\end{table*}%

\begin{table*}[!htbp]
\fontsize{0.18cm}{0.16cm}\selectfont
\caption{Reviewed research articles from 2010 to 2013 classified considering the \textbf{cloud branches} of the proposed taxonomy.}
\label{tab:surveyPreCloud1}
\begin{center}
\setlength\tabcolsep{4pt}
    \begin{tabular}{l|cccccc|ccccc|cc|cc|cccc|cccccc|}
    \toprule
        & \multicolumn{25}{c}{Cloud taxonomy branches} \\[5pt]
        & \rot{S/T/.../C/SaaS} & \rot{S/T/.../C/PaaS} & \rot{S/T/.../C/IaaS/Public} & \rot{S/T/.../C/IaaS/Private} &  \rot{S/T/.../C/IaaS/Hybrid} & \rot{S/T/.../C/IaaS/Community}
        & \rot{S/OBJ/P/MP/Min. DC costs} & \rot{S/OBJ/P/MP/Min. ex. costs} & \rot{S/OBJ/P/Max. user QoS} & \rot{S/OBJ/C/Min. ex. costs} & \rot{S/OBJ/C/Max. QoS} 
        & \rot{S/I/S/MC/Charging Model} & \rot{S/I/S/MC/Price} &  \rot{S/A/V/Virtual machines}  & \rot{S/A/V/Containers} &  \rot{S/I/S/VMs/Security}  & \rot{S/I/S/VMs/P/Processor} & \rot{S/I/S/VMs/P/Disk} & \rot{S/I/S/VMs/P/Network} 
        & \rot{S/L/P/SaaS} & \rot{S/L/P/PaaS} & \rot{S/L/P/IaaS} & \rot{S/L/C/User} & \rot{S/L/C/P/SaaS} & \rot{S/L/C/P/PaaS--Broker} \\
        \midrule
\rowcolor{Grey7}\textbf{2013} & & & & & & & & & & & & & & & & & & & & & & & & & \\
Abrishami, S. et al. \cite{Abrishami2013158}  & $\circ$ & $\circ$ & $\bullet$ & $\circ$ & $\circ$ & $\circ$ & $\circ$ & $\circ$ & $\circ$ & $\bullet$ & $\bullet$ & $\circ$ & $\bullet$ & $\circ$ & $\circ$ & $\circ$ & $\bullet$ & $\circ$ & $\circ$ & $\circ$ & $\circ$ & $\circ$ & $\bullet$ & $\circ$ & $\bullet$ \\
\rowcolor{Grey9} Dhillon, J. S. et al. \cite{Dhillon2013} & $\circ$ & $\circ$ & $\bullet$ & $\bullet$ & $\circ$ & $\circ$ & $\circ$ & $\circ$ & $\bullet$ & $\circ$ & $\circ$ & $\circ$ & $\circ$ & $\bullet$ & $\circ$ & $\circ$ & $\circ$ & $\circ$ & $\circ$ & $\circ$ & $\circ$ & $\bullet$ & $\circ$ & $\circ$ & $\circ$ \\
Ergu, D. et al. \cite{Ergu2013} & $\circ$ & $\circ$ & $\circ$ & $\bullet$ & $\circ$ & $\circ$ & $\circ$ & $\circ$ & $\bullet$ & $\circ$ & $\circ$ & $\circ$ & $\circ$ & $\circ$ & $\circ$ & $\circ$ & $\circ$ & $\circ$ & $\circ$ & $\bullet$ & $\bullet$ & $\circ$ & $\circ$ & $\circ$ & $\circ$ \\
\bottomrule
\end{tabular}
\end{center}
\label{tab3}
\end{table*}%

\begin{table*}[!htbp]
\fontsize{0.18cm}{0.16cm}\selectfont
\caption{Reviewed research articles from 2010 to 2013 classified considering the \textbf{pre-cloud branches} of the proposed taxonomy.}
\label{tab:surveyPreCloud2}
\begin{center}
\setlength\tabcolsep{4pt}
    \begin{tabular}{l|ccc|ccc|ccc|cccc|cccc|ccc|cc|}
    \toprule
        & \multicolumn{22}{c}{Pre-cloud taxonomy branches} \\[5pt]
        & \rot{S/O/Centralized} & \rot{S/O/Decentralized} & \rot{S/O/Hierarchical} & \rot{S/IA/Uncertain} &  \rot{S/I/A/Stochastic} & \rot{S/I/A/Precise}
        & \rot{S/I/S/Resources} & \rot{S/I/S/Jobs} & \rot{S/I/S/Users} & \rot{S/F/Online} & \rot{S/F/Offline/Static} 
        & \rot{S/F/Offline/Semi-Static}  & \rot{S/F/Offline/Dynamic} & \rot{S/A/T/Independent} & \rot{S/A/T/Dependent} & \rot{S/A/S/Independent} & \rot{S/A/S/Dependent}
        & \rot{S/OPT/Single criterium} &  \rot{S/OPT/Bi-criteria} &  \rot{S/OPT/Multi-criteria} & \rot{S/OUT/Imprecise} & \rot{S/OUT/Exact} \\ 
         \midrule
\rowcolor{Grey7} \textbf{2013} & & & & & & & & & & & & & & & & & & & & & & \\
Abrishami, S. et al. \cite{Abrishami2013158} & $\bullet$	& $\circ$	& $\circ$	& $\circ$	& $\circ$	& $\bullet$	& $\bullet$	& $\bullet$	& $\circ$	& $\circ$	& $\bullet$	& $\circ$	& $\circ$	& $\circ$	& $\bullet$	& $\circ$	& $\circ$	& $\circ$	& $\bullet$	& $\circ$	& $\circ$	& $\bullet$\\
\rowcolor{Grey9} Dhillon, J. S. et al. \cite{Dhillon2013}  & $\circ$	& $\bullet$	& $\circ$	& $\circ$	& $\circ$	& $\bullet$	& $\bullet$	& $\circ$	& $\circ$	& $\circ$	& $\bullet$	& $\circ$	& $\circ$	& $\circ$	& $\circ$	& $\circ$	& $\circ$	& $\bullet$	& $\circ$	& $\circ$	& $\circ$	& $\bullet$\\
Ergu, D. et al. \cite{Ergu2013}  & $\bullet$	& $\circ$	& $\circ$	& $\circ$	& $\circ$	& $\circ$	& $\bullet$	& $\bullet$	& $\circ$	& $\circ$	& $\bullet$	& $\circ$	& $\circ$	& $\bullet$	& $\circ$	& $\circ$	& $\circ$	& $\circ$	& $\circ$	& $\bullet$	& $\circ$	& $\bullet$\\
\rowcolor{Grey9}   Genez, T. et al. \cite{Genez2013}& $\bullet$	& $\circ$	& $\circ$	& $\circ$	& $\circ$	& $\bullet$	& $\bullet$	& $\bullet$	& $\circ$	& $\circ$	& $\bullet$	& $\circ$	& $\circ$	& $\circ$	& $\bullet$	& $\circ$	& $\circ$	& $\circ$	& $\bullet$	& $\circ$	& $\circ$	& $\bullet$\\
\bottomrule
\end{tabular}
\end{center}
\label{tab4}
\end{table*}%

	From Tables \ref{tab:surveyCloud1}, \ref{tab:surveyCloud2}, \ref{tab:surveyPreCloud1}, and \ref{tab:surveyPreCloud2}, we highlight the following conclusions based on the classification resulted from the proposed taxonomy:
	\paragraph{Organization} the scheduling problem is commonly approached from a centralized perspective, where a central scheduler has information about the system as a whole. 
	\paragraph{Input Accuracy} In the input \textit{accuracy} taxonomy, most scheduling work assumes data input is accurate, while only a few papers consider stochastic or uncertain inputs.  Between $2014$ and $2016$, works considering uncertainty in the input data received more attention than in previous years.
	\paragraph{Input Semantics} pre-cloud \textit{semantics} of the input data highlights the established assumption that schedulers often consider as input information about the computing resources as well as about the jobs to be run. Information about the user is less common, with a few works considering budget, for instance.
	\paragraph{Frequency} The most common scheduling frequency for algorithms is offline/static. A few works study online schedulers, and some works study semi-static or dynamic schedulers. 
	\paragraph{Application} Regarding the application model considered by the scheduler in the pre-cloud taxonomy, most works focus on tasks, both dependent and independent. However, a good amount of research is also dedicated to services, with most of them focusing on workflows (dependent services).
	\paragraph{Optimization} Scheduling research in computing, for a long time, focused on a single criteria, often makespan. In this review of recent literature, cloud computing scheduling brought the focus to bi- and multi-criteria scheduling approaches, where costs and resource utilization play an important role.
	\paragraph{Output} As a consequence of the common assumption that input data is precise, the scheduler output is more commonly considered to be also exact, with imprecise outputs receiving a little bit more attention in the last years.
	\paragraph{Target} IaaS clouds are the most common target system considered by scheduling research in clouds. As IaaS is often a basis of PaaS and SaaS, one could expect this focus on IaaS. On the other hand, PaaS and SaaS also need higher-level schedulers that can better understand application requirements.
	\paragraph{Location}   Literature often considers the cloud client itself as the host of the scheduler, where information about application and users are used to take the best scheduling decision. When the scheduler is assumed to run in the provider, a broker is commonly the system assumed to be hosting its decision-making process (and having an IaaS as the target system). IaaS clouds are also often considered as hosting the scheduler, even though more recently authors considered it fewer times than in cloud computing's earlier years. 
	\paragraph{Input Semantics Cloud} Many schedulers in cloud assume they have information about the VMs capacities, specially processing and network. Also, as an important cloud aspect, monetary costs (mostly price) is also a common input data for schedulers in cloud.
	\paragraph{Objective} As many works consider the scheduler to be in the client, the objectives considered by the scheduler optimization in cloud research take into account the client point of view: application execution costs and applications QoS. Nevertheless, few works have also taken into account the provider's point of view by considering data-center and execution costs.
	\paragraph{Application Cloud} Virtual machines and containers were introduced recently, and schedulers in clouds can consider them as the ``application'' to be scheduler. Only a few works have considered virtual machines or containers scheduling, with most of them being dedicated to VM allocation in cloud data centers.

 \section{Future directions}
 \label{sec:chal}
	Schedulers will need to keep evolving along with distributed systems to properly account new characteristics that arise~\cite{StephenSolved}. In this section we highlight scheduling challenges that still exist in current distributed system, supporting them with information from the literature review above, as well as devise problems that will be important to be addressed in the future of distributed systems. The discussion presented here is supported by data presented as supplementary material for this article.
	
\paragraph{Input data} One of the main practical challenges that are present in job scheduling in computer systems is regarding the quality of the input data. In practice, although some applications can reliably estimate job execution time from past executions, for many other applications jobs running times are either unknown or a rough estimation, leading to imprecision in the schedule results. In pre-cloud systems, this can result in problems such as delayed application execution, increased makespan, deadline misses, among others. In clouds, an additional direct consequence of such unhandled uncertainty is increased costs. Uncertainties can be a result of no previous knowledge about jobs and resources, but also from resource contention when concurrency is present. In the latter, network contention is of special importance when data dependencies and data transfers exist among jobs of the same user or from users sharing the same infrastructure. Most schedulers do not take resource contention into consideration, what can actually bring more uncertainties to the application execution times. Schedulers that are robust to such uncertainties are necessary to avoid  the discussed pitfalls and to increase reliability on schedulers estimates.

 \paragraph{QoS and QoE} Many schedulers take into account multiple criteria in the decision making. These criteria are commonly associated with quality of service metrics, which in turn are associated with a higher level requirement from the user. To allow a scheduler to autonomously adapt itself, it must understand those higher level requirements, referred as quality of experience (QoE). This involves translating the user's input data (requirements) into lower quality of service levels that can be monitored and adapted according to the user quality of experience. Currently, QoS is challenging for schedulers due to variable requirements from different applications, and this challenge becomes even greater when QoS metrics observed are the same, but a different perception of quality can come from different users for the same application.
	
\paragraph{Edge/Fog computing} As cloud computing became more mature, the emergence of models to cover requirements not fulfilled by clouds also evolved. One promising distributed system architecture evolution is bringing computing capacity to the edges of the network to reduce latencies \cite{Bittencourt2016}, as for example in Fog Computing \cite{bonomi2012fog}. As computing capacity is moved closer to the user, but cloud computing do still exists to provide larger capacity, a hierarchical system emerges, with each level having its constraints and capabilities. For instance, in Fog Computing \textit{Cloudlets} closer to the user can fulfil lower-latency requirements and locally optimized decisions, while cloud computing keeps having its role as a centralized, high-capacity computing facility. This hierarchical heterogeneous system introduces new variables to be considered by the schedulers, with latency becoming a relevant optimization criterium.
	
\paragraph{Mobility} Mobile applications are more and more complex as smart devices become ubiquitous, with users often carrying more than one (sometimes multiple with wearable computing) processing-capable, networked devices. Such applications often rely on data storage and processing capacities from the cloud to provide relevant functionalities to mobile users. In this context, mobile users have at a given time and location a set of data and processing tasks that is a subset of all his/her needs. Edge computing, e.g. Fog Computing, associated with mobility can be used to maintain these subsets closer to the user. This is  desirable to reduce network traffic and associated delays and latencies \cite{Bittencourt2015, Bittencourt2017}. To make this effective, resource allocation and scheduling play an important role, and modelling the scheduling problem introducing time and mobility variables is a challenge that arises. These models would have to consider user positioning and movement predictions in order to be able to move data in advance and reduce user's perception of latency in a variety of mobility scenarios and edge resource availability.
	
\paragraph{Service levels} Schedulers are usually developed for specific service levels. The reviewed literature presents an imbalance in studies for IaaS and other service levels, which suggests schedulers that can take advantage of different service levels are yet to be developed. Moreover, studies to understand the relation between costs and resource utilization at the different cloud service levels could help in reducing cloud costs from the user perspective.
	
\paragraph{Charging models} Besides being service-level-specific, cloud offerings also bring a variety of charging models with different charging intervals. Notwithstanding, most schedulers do not take the charging models into account. Two traditional charging models for virtual machines are \textit{on-demand} and \textit{reserved} instances. Auction-style instances with variable prices also exist. While some applications can have strict requirements that better match with a charging model that offers a more reliable instance, other applications might be prone to failures, where cheaper instances could be useful. A comprehensive scheduling model that captures the relation between charging models, charging intervals, and pricing would help in establishing lower execution costs within the applications QoS requirements. 

\paragraph{IoT and BigData} Internet of Things and Big Data challenges are also relevant to scheduling in distributed systems. With predictions on the number of devices connected to the Internet reaching 50-100 billion \cite{perera2014context}, the amount of data produced, transferred, and processed will also increase unprecedentedly. Raw data generated by the plethora of connected devices can be stored for further processing or future access, or it can be processed before storage, being reduced to more condensed, meaningful data. The on-the-fly processing of such stream of data is referred as \textit{stream processing} or \textit{complex event processing}. In this kind of processing, incoming data passes through a set of operators to be filtered/aggregated and stored. The way these operators are allocated to the available computing resources have impact in the delay observed between when the data input has taken place and its resulting computation result is obtained. With billions of devices data streams realizing a set of operations, the scale of the scheduling problem can make current decision-making methods unfeasible. Developing fast, online scheduling that can handle big data is a challenging issue that can impact the future of the Internet of Things. Online schedulers are much less common in the past literature than offline schedulers, which makes the development of effective fast online scheduling even more challenging.
	
\paragraph{Robustness} Models that capture the relation between scheduling and resilience, specially for large applications, can reduce execution times and costs. Any changes in the execution of an application, such as changes in application topology, resources configuration, software configuration, input data, or data sources can turn a successful execution into a failure. An application scheduler that is able to capture failure occurrences from a resilience model and avoid allocations that can potentially fail is a challenging research topic which results can benefit the execution of large applications by avoiding interruption, and consequent waste of money and time.
	
\paragraph{Online optimization} Most schedulers assume the sequence of jobs is known a-priori. Online scheduling algorithms that capture many characteristics of tasks and systems (such offline schedulers do) are still a challenge in distributed systems. One of the difficulties is that previously dispatched jobs might need to be relocated in order to achieve the aim of a given objective function. Process migration can be costly and complicated, but with the adoption of virtualization (VMs and containers), migration is feasible. Online schedulers that include migration costs and the impact of migration in the running applications can bring online schedulers to attention, with more comprehensive online schedulers becoming relevant.
	
\paragraph{VM placement} Clouds provide, through virtual machines, resource sharing among tenants. Although resource sharing through VMs promotes software isolation improves resource utilization, it does no guarantee performance isolation. Virtual machine allocation, or virtual machine placement, algorithms that can avoid performance interference among cloud stakeholders are a challenge because of lack of information or knowledge about the applications users are going to deploy. Moreover, a trade-off between performance isolation and cost is present: a guarantee of performance isolation can actually mean less users/applications per host, reducing resource utilization and consequently increasing infrastructure costs. Application profiling can be used to improve input information for the VM allocation algorithms, but care must be taken not to violate users privacy. 

\paragraph{Security and privacy} With the resource sharing and transparent data/processing offloading promoted by virtualization, security and privacy are increasing concerns. As the user cannot control with whom his VM will share the physical resources, the co-located VMs can potentially be owned by anyone around the globe. Although security is a general concern on networked environments, resource allocation and scheduling can take it into account to reduce risks according to application and users requirements. This introduces new constraints to the scheduling problem, what can again result in poorer resource utilization and increased costs.

\section{Conclusion}
\label{sec:con}
	The classification of scheduling solutions is challenging due to the large amount of literature as well as the number of variations of the problem. These very same reasons make the identification of relevant topics in the scheduling problem an important task to be regularly performed by both young and experienced researchers in the field.

	In this paper we review several aspects of the scheduling literature for distributed systems, and proposed a taxonomy that encompasses the scheduler organization in the system, the scheduler input and output data, the frequency the scheduler runs, the application model and target system, and the scheduling objectives. The taxonomy is extended to include branches that appeared with the more recent cloud computing paradigm, bringing new target system aspects, the utility perspectives in terms of target system and scheduling objectives, as well as different application model resulted from the cloud virtualized characteristic.
	
	To corroborate the taxonomy, we reviewed a total of $114$ papers from the scheduling literature, identifying research according to the proposed taxonomy branches. This resulted in an analysis of the latest scheduling research in distributed systems from a variety of aspects. As a result from the literature review, we also discussed trends and challenges that can be focus of research in the upcoming years.

	The work of classifying scheduling research does not come to a foreseeable end. The need for classification and survey of scheduling literature for distributed systems will remain in the future, especially considering the internet of things, mobility, and big data resulting from the ever increasing number of connected devices, as well as the continued existence of computing as a utility service.

\section*{Acknowledgements}
	This work was partially funded by the following agencies/grants: CAPES, CNPq, and  grants \#2009/15008-1 and \#2015/16332-8, S\~{a}o Paulo Research Foundation (FAPESP).

%

\end{document}